\documentclass{sig-alternate-2013}

\newfont{\mycrnotice}{ptmr8t at 7pt}
\newfont{\myconfname}{ptmri8t at 7pt}
\let\crnotice\mycrnotice%
\let\confname\myconfname%

\usepackage{amsmath}
\usepackage{latexsym}
\usepackage{amssymb}
\usepackage{verbatim}
\usepackage{fancyvrb}
\usepackage{todonotes}
\usepackage{algorithm}
\usepackage{algpseudocode}
\usepackage{url}
\usepackage{fixltx2e}
\usepackage{xspace}
\usepackage{xfrac}
\usepackage{rotating}
\usepackage{booktabs}
\usepackage{paralist}
\usepackage{comment} 
\usepackage{graphicx}
\usepackage{caption}
\usepackage{subcaption}
\usepackage{tabularx}
\usepackage{epstopdf}

\algnotext{EndFor}
\algnotext{EndIf}
\algnotext{EndProcedure}
\algnotext{EndFunction}

\newcommand{\mytitle}{LINVIEW: Incremental View Maintenance\\ for Complex Analytical Queries}
\title{\mytitle\thanks{This work was supported by ERC Grant 279804.}}

\numberofauthors{1}
\author{
\alignauthor Milos Nikolic ~~ Mohammed ElSeidy ~~ Christoph Koch \\[1ex]
       \affaddr{\{milos.nikolic, mohammed.elseidy, christoph.koch\}@epfl.ch}\\
       \affaddr{\'Ecole Polytechnique F\'ed\'erale de Lausanne}
}

\newtheorem{theorem}{Theorem}[section]
\newtheorem{example}[theorem]{Example}
\newtheorem{definition}[theorem]{Definition}
\def\punto{$\hspace*{\fill}\Box$}

\newcommand{\tuple}[1]{\langle#1\rangle}
\newcommand{\pluseq}{\mathrel{+}=}
\newcommand{\tr}[1]{#1^{\text{T}}}
\newcommand{\inv}[1]{#1^{-1}}
\def\mul{\,}
\def\mulvb{\(\,\)}

\def\du{u}
\def\dv{v}
\def\duA{\du_{\text{A}}}
\def\dvA{\dv_{\text{A}}}

\def\dU{U}
\def\dV{V}
\def\dQ{Q}
\def\dR{R}
\def\dW{W}
\def\dZ{Z}
\def\dUB{\dU_{\text{B}}}
\def\dVB{\dV_{\text{B}}}
\def\Du{\Delta_{\text{A}}}
\def\Deltavb{\(\Delta\)}
\def\bigO{\mathcal{O}}

\def\system{\textsc{Linview}\xspace}

\def\incrLin{\textsc{IncrLin}\xspace}

\def\reevalExp{\textsc{ReevalExp}\xspace}
\def\incrExp{\textsc{IncrExp}\xspace}

\newcommand{\incrSkips}[1]{\textsc{IncrSkip-#1}\xspace}

\newcommand{\smartparagraph}[1]{\vspace{1ex}\noindent{\bf #1}}

\begin{document}

\maketitle

\begin{abstract}
Many analytics tasks and machine learning problems
can be naturally expressed by iterative linear algebra
programs. In this paper, we study the incremental view maintenance
problem for such complex analytical queries.
We develop a framework, called \system, for capturing deltas of linear algebra
programs and understanding their computational cost. Linear algebra operations tend to
cause an avalanche effect  
where even very local changes to the input matrices spread out and infect all
of the intermediate results and the final view, causing incremental view
maintenance to lose its performance benefit over re-evaluation.
We develop techniques based on matrix
factorizations to contain such epidemics of
change. As a consequence, our techniques make incremental view maintenance
of linear algebra practical and usually substantially cheaper than
re-evaluation.
We show, both analytically and experimentally, the usefulness
of these techniques 
%through our compiler framework 
when applied to standard analytics tasks. 
Our evaluation demonstrates the efficiency of \system in generating parallel
incremental programs that outperform re-evaluation techniques by more than an order of magnitude.
\end{abstract}

%\noindent{\bf Categories and Subject Descriptors: } H.2 [Database
%Management]: Systems;

%\noindent{\bf Keywords}: linear algebra; incremental view maintenance;
%compilation; machine learning; Spark

%\todo[inline]{Refer to incremental maintenance instead of incremental evaluation}
\section{Introduction}

Linear algebra plays a major role in modern data analysis. 
Many state-of-the-art data mining and machine learning algorithms 
are essentially linear transformations of vectors and matrices, 
often expressed in the form of iterative computation.
Data practitioners, engineers, and scientists utilize such algorithms 
to gain insights about the collected data.

Data processing has become increasingly expensive in the era of big data.
Computational problems in many application domains, like social graph analysis,
web analytics, and scientific simulations, often have to process petabytes of
multidimensional array data.
Popular statistical environments, like R and MATLAB, offer
high-level abstractions that simplify programming but lack the support for
scalable full-dataset analytics.
Recently, many scalable frameworks for data analysis have emerged:
MADlib~\cite{MADlib_PVLDB2012} and
%Bismarck~\cite{Bismarck_Sigmod2012}
Columbus~\cite{Columbus_Sigmod2014} 
for in-database scalable analytics,
SciDB~\cite{SciDB_CIDR2009}, SciQL~\cite{SciQL_Sigmod2013},
and RasDaMan~\cite{RasDaMan_Sigmod1998} for in-database array processing,
Mahout and MLbase~\cite{DBLP:conf/cidr/KraskaTDGFJ13} for
machine learning and data mining on Hadoop and
Spark~\cite{Zaharia2010}.
High-performance computing relies on optimized libraries, like Intel
MKL and ScaLAPACK, to accelerate the performance of
matrix operations in data-intensive computations.
All these solutions primarily focus on efficiently processing large volumes of
data.

Modern applications have to deal with not just big, but also rapidly changing
datasets.
A broad range of examples including clickstream analysis, algorithmic trading,
network monitoring, and recommendation systems compute realtime analytics
over streams of continuously arriving data.
Online and responsive analytics allow data miners, analysts, and statisticians
to promptly react to certain, potentially complex, conditions in the data; or to
gain preliminary insights from approximate or incomplete results at very early
stages of the computation.
Existing tools for large-scale data analysis often lack support for dynamic
datasets.
High data velocity forces application developers to build ad hoc solutions in
order to deliver high performance.
More than ever, data analysis requires efficient and scalable solutions to cope
with the ever-increasing volume and velocity of
% generated
data.

Most datasets evolve through changes that are small relative to the overall dataset size.
For example, the Internet activity of a single user, like a user's purchase history or movie ratings, represents only a tiny portion of the collected data. 
Recomputing data analytics on every (moderate) dataset change is far from efficient.

These observations motivate {\em incremental data analysis}.
Incremental processing combines the results of previous analyses with incoming
changes to provide a computationally cheap method for updating the result. The
underlying assumption that these changes are relatively small
allows us to avoid re-evaluation of expensive operations.
In the context of databases, incremental processing -- also known as incremental
view maintenance -- reduces the cost of query evaluation by simplifying or
eliminating join processing for changes in the base relations~\cite{Gupta1999,
DBToaster:VLDBJ2014}.
However, simulating multidimensional array computations on top of traditional
RDBMSs can result in poor performance~\cite{SciDB_CIDR2009}.
Data stream processing systems~\cite{Motwani03queryprocessing, Abadi2005} also
rely on incremental computation to reduce the work over finite windows of input
data.
Their usefulness is yet limited by their window semantics and inability to
handle long-lived data.

In this work we focus on incremental maintenance of analytical queries written
as (iterative) linear algebra programs.
A program consists of a sequence of statements performing operations on vectors
and matrices.
For each matrix (vector) that dynamically changes over time, we define a trigger
program describing how an incremental update affects the result of each
statement (materialized view).
A {\em delta expression} of one statement captures the difference between the
new and old result.
This paper shows how to propagate delta expressions through subsequent
statements while avoiding re-evaluation of computationally expensive operations,
like matrix multiplication or inversion.

\begin{figure}[t]
\centering
\vspace{-1mm}
\includegraphics[width=0.9\columnwidth]{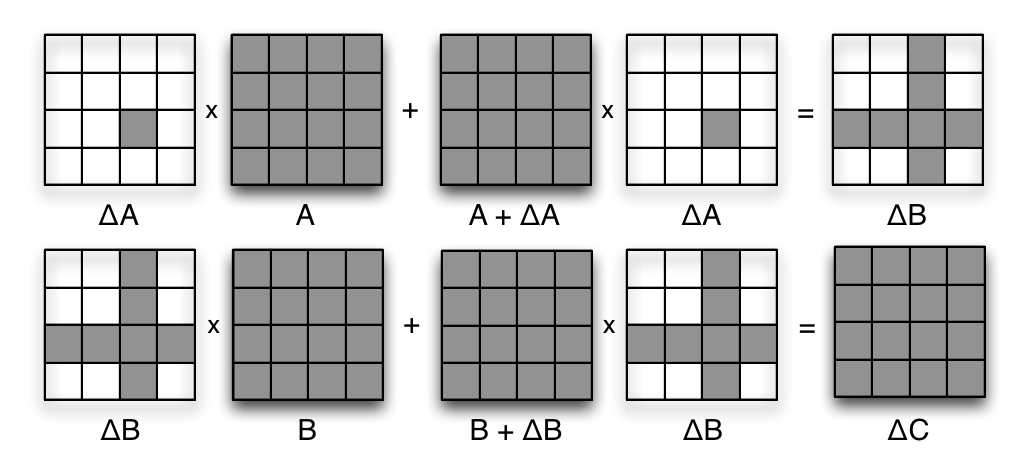}
\vspace{-3mm}
\caption{
%Graphical representation of the 
Evaluation of $\Delta{B}$ and $\Delta{C}$ for a single entry change in $A$. Gray entries have nonzero values.}
\label{fig:example_delta_evaluation}
\vspace{-3mm}
\end{figure}

\vspace{-2.5mm}
\begin{example} \label{ex:simple_delta}
\em
Consider the program that computes the fourth power of a given matrix $A$.
{
\vspace{-1.5mm}
\begin{Verbatim}[commandchars=\\\{\}]
B := A\mulvb{}A;
C := B\mulvb{}B;
\end{Verbatim}
\vspace{-1.5mm}
}
The goal is to maintain the result $C$ on every update of $A$ by $\Delta{A}$. 
We compare the time complexity of two computation strategies: re-evaluation and incremental maintenance.
The re-evaluation strategy first applies $\Delta{A}$ to $A$ and then performs 
two $\bigO(n^3)$\footnote{\scriptsize This example assumes the traditional
cubic-time bound for matrix multiplication. Section~\ref{sec:linear_programs}
generalizes this cost for asymptotically more efficient methods.} matrix
multiplications to update $C$.

The incremental approach exploits the associativity and distributivity of matrix multiplication to compute a delta expression for each statement of the program.
The trigger program for updates to $A$ is:
{
\vspace{-1mm}
\begin{Verbatim}[commandchars=\\\{\}]
ON UPDATE A BY \Deltavb{}A:
    \Deltavb{}B := (\Deltavb{}A)A + A(\Deltavb{}A) + (\Deltavb{}A)(\Deltavb{}A);
    \Deltavb{}C := (\Deltavb{}B)B + B(\Deltavb{}B) + (\Deltavb{}B)(\Deltavb{}B);  
    A += \Deltavb{}A;  B += \Deltavb{}B;  C += \Deltavb{}C;
\end{Verbatim}
\vspace{-1mm}
} 
Let us assume that $\Delta{A}$ represent a change of one cell in $A$. 
Fig.~\ref{fig:example_delta_evaluation} shows the effect of that change on
$\Delta{B}$ and $\Delta{C}$\,\footnote{\scriptsize For brevity, 
we factor the last two monomials in $\Delta{B}$ and $\Delta{C}$.}.
The shaded regions represent entries with nonzero values. 
The incremental approach capitalizes on the sparsity of $\Delta{A}$ and $\Delta{B}$ to compute $\Delta{B}$ and $\Delta{C}$ in $\bigO(n)$ and $\bigO(n^2)$ operations, respectively. 
Together with the cost of updating $B$ and $C$, incremental evaluation of the
program requires $\bigO(n^2)$ operations, clearly cheaper than re-execution. \punto
\end{example}

\vspace{-1.5mm}
The main challenge in incremental linear algebra is how to represent and
propagate delta expressions.
Even a small change in a matrix (e.g., a change of one entry) might have an
avalanche effect that updates every entry of the result matrix.
In Example~\ref{ex:simple_delta}, a single entry change in $A$ causes changes of
one row and column in $B$, which in turn pollute the entire matrix $C$.
If we were to propagate $\Delta{C}$ to a subsequent expression -- for example,
the expression $D=C \mul C$ that computes $A^8$ -- then evaluating $\Delta{D}$
would require two full $\bigO(n^3)$ matrix multiplications, which is obviously more expensive than
recomputing $D$ using the new value of $C$.

To confine the effect of such changes and allow efficient evaluation, 
we represent delta expressions in a {\em factored form}, as products of low-rank matrices.
For instance, we could represent $\Delta{B}$ as a vector outer product for single entry updates in $A$. 
Due to the associativity and distributivity of matrix multiplication, 
the factored form allows us to choose the evaluation order that completely avoids expensive matrix multiplications.

To the best of our knowledge, this is the first work done towards efficient
incremental computation for large scale linear algebra data
analysis programs. In brief, our contribution can be summarized as follows:
% \begin{enumerate} \addtolength{\topsep}{-0.3ex}
% \addtolength{\labelsep}{-0.3ex} \addtolength{\itemsep}{-1ex}

\smartparagraph{1.}{ We present a framework for incremental maintenance of
linear algebra programs that:
\begin{inparaenum}[\itshape a\upshape)]
\item represents delta expressions in compact factored forms that confine the avalanche effect of input changes, as seen in Example~\ref{ex:simple_delta}, and thus cost.
\item utilizes a set of transformation rules that metamorphose linear algebra programs into their cheap functional-equivalents that are optimized for dynamic datasets.
\end{inparaenum}
}

\smartparagraph{2.}{ We demonstrate analytically and experimentally the efficiency of incremental processing on various fundamental data analysis methods including ordinary least squares, batch gradient descent, PageRank, and matrix powers.}

\smartparagraph{3.}{ We have built \system, a compiler for incremental data
analysis that exploits these novel techniques to generate efficient update
triggers optimized for dynamic datasets. The compiler is easily extensible to couple with
any underlying system that supports matrix manipulation primitives.
%,e.g., BLAS. 
We evaluate the performance of \system's generated code over two
different platforms:
\begin{inparaenum}[\itshape a\upshape)] \item Octave programs running
on a single machine and \item parallel Spark programs running over a
large cluster of Amazon EC2 nodes.
\end{inparaenum}
 Our results show that incremental evaluation provides an order of magnitude
 performance benefit over traditional re-evaluation.}

%\smartparagraph{3.}{ We have built an extensible system for scalable data analysis that exploits the techniques discussed above to generate efficient update triggers optimized for execution on parallel processing platforms such as Spark~\cite{Zaharia2010}.}

The paper is organized as follows: 
Section~\ref{sec:related} reviews the related work,
Section~\ref{sec:linear_programs} establishes the terminology and
computational models used in the paper,
Section~\ref{sec:incremental} describes how to compute, 
represent, and propagate delta expressions,
Section~\ref{sec:use_cases} analyzes the efficiency of incremental maintenance
on common data analytics, 
Section~\ref{sec:systems} gives the system overview, and 
Section~\ref{sec:experiments} experimentally validates our analysis.

\section{Related Work}
\label{sec:related}
This section presents related work in different directions.

\noindent {\bf IVM and Stream Processing.}\ Incremental View Maintenance
techniques \cite{Blakeley1986, DBToaster:VLDBJ2014, Gupta1999} support 
incremental updates of database materialized views by employing differential 
algorithms to re-evaluate the view expression. Chirkova \emph{et al.\
}\cite{Chirkova2012} present a detailed survey on this direction.
Data stream processing engines~\cite{Abadi2005, Motwani03queryprocessing,
Arasu2004} incrementally evaluate continuous queries as windows advance over
unbounded input streams. 
In contrast to all the previous, this paper targets incremental maintenance of
linear algebra programs as opposed to classical database (SQL) queries.
The linear algebra domain has different semantics and primitives,   
thus the challenges and optimization techniques differ widely.

\noindent {\bf Iterative Computation.} \ Designing frameworks and computation
models for iterative and incremental computation has received much attention
lately.
Differential dataflow~\cite{McSherry2013} represents a new model of incremental
computation for iterative algorithms, which relies on differences (deltas) being
smaller and computationally cheaper than the inputs.
This assumption does not hold for linear algebra programs because of the
avalanche effect of input changes.
Many systems optimize the MapReduce framework for iterative applications using
techniques that cache and index loop-invariant data on local disks and
persist materialized views between iterations~\cite{Bu2010,Ekanayake2010,Zhang2012}.
More general systems support iterative computation and the DAG execution model,
like Dryad~\cite{Isard2007} and Spark~\cite{Zaharia2010};
Mahout, MLbase~\cite{ DBLP:conf/cidr/KraskaTDGFJ13} and 
others~\cite{Das2010, Venkataraman2012, Mihaylov2012} provide scalable machine 
learning and data mining tools. All these systems are orthogonal to our work. 
This paper is concerned with the efficient re-evaluation of programs under 
incremental changes. Our framework, however, can be easily coupled with any of 
these underlying systems.

\noindent {\bf Scientific Databases.} \ There are also database systems
specialized in array processing. RasDaMan~\cite{RasDaMan_Sigmod1998}
and AML~\cite{Marathe2002} provide support for expressing and
optimizing queries over multidimensional arrays, but are not geared towards
scientific and numerical computing.
%Queries are translated into an array algebra and  optimized using a large collection of transformation rules.
%language for array manipulation alongside with a suite of query optimization techniques.
ASAP~\cite{Stonebraker2007} supports scientific computing primitives on
a storage manager optimized for storing multidimensional arrays.
RIOT~\cite{Zhang2009} also provides an efficient \emph{out-of-core}
framework for scientific computing. However, none of these systems
support incremental view maintenance for their workloads.

\noindent {\bf High Performance Computing.} \ The advent of numerical and
scientific computing has fueled the demand for efficient matrix manipulation
libraries.
BLAS~\cite{Dongarra1990} provides low-level routines representing common linear
algebra primitives to higher-level libraries, such as LINPACK,
LAPACK, and ScaLAPACK for parallel processing.
Hardware vendors such as Intel and AMD and code generators such
as ATLAS~\cite{Whaley1999} provide highly optimized BLAS
implementations.
In contrast, we focus on incremental maintenance of programs through
efficient transformations and materialized views. The \system compiler
translates expensive BLAS routines to cheaper ones and thus
further facilitates adoption of the optimized implementations.

\noindent {\bf PageRank.} \ There is a huge body of literature that
is focused on PageRank, including the Markov chain model, solution
methods, sensitivity and conditioning, and the updating problem. Surveys can
be found in~\cite{Langville2004, Berkhin2005}. The updating problem studies the
effect of perturbations on the Markov chain and PageRank
models, including sensitivity analysis, approximation, and exact evaluation
methods. In principle, these methods are particularly tailored for these
specific models. In contrast, this paper presents a novel model and
framework for efficient incremental evaluation of \emph{general} linear algebra
programs through domain specific compiler translations and efficient code
generation.

\noindent {\bf Incremental Statistical Frameworks.} \ Bayesian
inference~\cite{Berger1985} uses Bayes' rule
to update the probability estimate for a hypothesis as
additional evidence is acquired. These frameworks support a variety of
applications such as pattern recognition and classification. 
Our work focuses on incrementalizing applications that can be
expressed as linear algebra programs and generating efficient incremental
programs for different runtime environments.

\noindent {\bf Programming Languages.}\ The programming language community has
extensively studied incremental computation and information
flow~\cite{Chen2012}. It has developed languages and compilation techniques for
translating high-level programs into executables that respond efficiently to
dynamic changes. \emph{Self-adjusting computation} targets incremental
computation by exploiting dynamic dependency graphs and change propagation
algorithms~\cite{Acar2009, Chen2012}. 
%that utilize a particular form of memoization techniques~\cite{Acar2009,
% Chen2012}.
These approaches: \begin{inparaenum}[\itshape a\upshape)] \item serve for 
general purpose programs as opposed to our domain specific approach, \item 
require serious programmer involvement by annotating modifiable portions of the
program, and \item fail to efficiently capture the propagation of deltas 
among statements as presented in this paper.
\end{inparaenum}

\section{Linear Algebra Programs}
\label{sec:linear_programs}

%In this section we introduce the notation and terminology used throughout the paper. 

%\subsection{Linear Algebra Programs}
Linear algebra programs express computations using vectors and matrices as high-level abstractions. 
The language used to form such programs consists of the standard matrix manipulation primitives: matrix addition, subtraction, multiplication (including scalar, matrix-vector, and matrix-matrix multiplication), transpose, and inverse.
A program expresses a computation as a sequence of statements, each consisting of an expression and a variable (matrix) storing its result.
The program evaluates these expressions on a given dataset of {\em input matrices} 
and produces the result in one or more {\em output matrices}. 
The remaining matrices are auxiliary program matrices; 
they can be manipulated (materialized or removed) for performance reasons.
For instance, the program of Example~\ref{ex:simple_delta} consists of two statements evaluating the expressions over an input matrix $A$ and an auxiliary matrix $B$. The output matrix $C$ stores the computation result.

\begin{comment}
Following the distributivity of matrix multiplication over addition, 
we can transform every linear algebra expression into a polynomial 
form, that is, represent it as a sum of monomials. 
Since the summation order of the monomials is irrelevant, we represent one expression 
as a set of monomials, with implicit summation among them.
We use $|f|$ to denote the number of monomials in expression $f$.
\end{comment}

\begin{comment}
Program $\mathcal{P}$ also indicates the matrices that can be changed, so called 
input matrices, and their update frequencies, denoted by $f_A$ for matrix A.
\end{comment}

\noindent{\bf Computational complexity}.\ In this paper, we refer to the cost of
matrix multiplication as $\bigO(n^\gamma)$ where $2\leq\gamma\leq3$.
In practice, the complexity of matrix multiplication, e.g., BLAS
implementations~\cite{Whaley1999}, is bounded by cubic $\bigO(n^3)$ time.
Better algorithms with an exponent of $2.37+\epsilon$ are known
(Coppersmith-Winograd and its successors); however, these algorithms are only
relevant for astronomically large matrices. Our incremental techniques remain
relevant as long as matrix multiplication stays asymptotically worse than
quadratic time (a bound that has been conjectured to be
achievable~\cite{Cohn2005}, but still seems far off). Note that the asymptotic
lower bound for matrix multiplication is $\Omega(n^2)$ operations because it
needs to process at least $2n^2$ entries.

\subsection{Iterative Programs}

Many computational problems are iterative in nature.
Iterative programs start from an approximate answer. Each iteration step
improves the accuracy of the solution until the estimated error drops below a
specified threshold.
Iterative methods are often the only choice for problems for which direct
solutions are either unknown (e.g., nonlinear equations) or prohibitively
expensive to compute (e.g., due to large problem dimensions).

In this work we study (iterative) linear algebra programs from the viewpoint
of incremental view maintenance (IVM).
The execution of an iterative program generates a sequence of results, one for each iteration step.
When the underlying data changes, IVM updates these results rather than re-evaluating them from scratch. 
We do so by propagating the delta expression of one iteration to subsequent iterations.
With our incremental techniques, such delta expressions are cheaper to evaluate
than the original expressions.

We consider iterative programs that execute a fixed number of iteration steps.
The reason for this decision is that programs using convergence thresholds might
yield a varying number of iteration steps after each update.
Having different numbers of outcomes per update would require incremental
maintenance to deal with outdated or missing old results; we leave
this topic for future work. 
By fixing the number of iterations, we provide a fair comparison of the
incremental and re-evaluation strategies\footnote{\scriptsize If the solution does not converge after a given
number of iterations, we can always re-evaluate additional steps.
% Exploring such hybrid evaluation strategies is out of the scope of the paper.
}.

\subsection{Iterative Models}

An iterative computation is governed by an iterative function that describes the
computation at each step in terms of the results of previous iterations
(materialized views) and a set of input matrices.
Multiple iterative functions, or \emph{iterative models}, might express the same
computation but by using different numbers of iteration steps.
For instance, the computation of the $k^{th}$ power of a matrix can be done in 
$k$ iterations or in $\log_{2}k$ iterations using the exponentiation by squaring method.
Each iterative model of computation comes with its own complexity.

Our analysis of iterative programs considers three alternative models that require different numbers of iteration steps to compute the final result.
These models allow us to explore trade-offs between computation time and memory consumption for both re-evaluation and incremental maintenance. 
%We present these models next. 

%\vspace{-1.0mm}
\smartparagraph{Linear Model.}{ The linear iterative model evaluates the result
of the current iteration based on the result of the previous iteration and a set
of input matrices $\mathcal{I}$. It takes $k$ iteration steps to compute $T_k$.
\vspace{-0.75mm}
\begin{align*}
T_i = \left\{
\begin{array}{l@{~}l@{~}l}
  f(\mathcal{I}) & \quad \quad & \mbox{for }\, i = 1 \\
  g(T_{i-1}, \mathcal{I}) & \quad \quad  & \mbox{for }\, i = 2, 3, \ldots
\end{array}
\right. %, \mbox{ for } i = 1, 2, \ldots k
\end{align*}
\vspace{-2.75mm}
}

\noindent{Example $A^k$: $T_1=A$ and $T_i=T_{i-1} \mul A$ for $2 \le i \le
k$.}

%\vspace{-1.0mm}
\smartparagraph{Exponential Model.}{ In the exponential model, the result of the
$i^{th}$ iteration depends on the result of the $(i/2)^{th}$ iteration.
The model makes progressively larger steps between computed iterations,
forming the sequence $T_1$, $T_2$, $T_4$, \ldots. It takes $\bigO(\log{k})$
iteration steps to compute $T_k$.
\vspace{-0.75mm}
\begin{align*}  
T_{i} = \left\{
\begin{array}{l@{~}l@{~}l}
  f(\mathcal{I}) & \quad \quad & \mbox{for }\, i = 1 \\
  g(T_{\sfrac{i}{2}}, \mathcal{I}) & \quad \quad & \mbox{for }\, i = 2, 4, 8
  \ldots
\end{array}
\right. %, \mbox{ for } i = 1, 2, 4,  \ldots k
\end{align*}
\vspace{-2.75mm}
}

\noindent{Example $A^k$: $T_1=A$ and $T_i=T_{\sfrac{i}{2}} \mul
T_{\sfrac{i}{2}}$ for $i=2,4,8,\ldots, k$.}

% \vspace{-1.0mm}
\smartparagraph{Skip Model.}{ Depending on the dimensions of input matrices,
incremental evaluation using the above models might be suboptimal costwise. The
skip-$s$ model represents a sweet spot between these two models.
For a given skip size $s$, it relies on the exponential model to compute $T_s$
(generating the sequence $T_2$, $T_4$, \ldots, $T_s$) and then generalizes the
linear model to compute every $s^{th}$ iteration (generating the sequence
$T_{2s}$, $T_{3s}$, \ldots).
% It exploits the exponential model to compute $T_s$ and then generalizes the
% linear model to compute every $s^{th}$ iteration, where $s$ denotes the skip
% step size. The generated sequence consists of two parts $T_1$, $T_2$, $T_4$,
% \ldots, $T_s$ and $T_{2s}$, $T_{3s}$, \ldots.
\vspace{-1.0mm}
\begin{align*}  
T_{i} = \left\{
\begin{array}{l@{~}l@{~}l}
  f(\mathcal{I}) & \quad \quad & \mbox{for }\, i = 1 \\
  g(T_{i/2}, \mathcal{I}) & \quad \quad & \mbox{for }\, i = 2, 4, 8, \ldots s \\
  h(T_{i-s}, T_{s}, \mathcal{I}) & \quad \quad & \mbox{for }\, i = 2s, 3s, \dots
\end{array}
\right. %, \mbox{ for } i = s, 2s, \ldots k
\end{align*}
\vspace{-2.75mm}
}

\noindent{Example $A^k$: For $s = 8$, we have $T_1=A$, then
$T_i=T_{\sfrac{i}{2}} \mul T_{\sfrac{i}{2}}$ for $i=2,4,8$, and $T_i=T_{i-8}
\mul T_8$ for $i=16, 24, 32, \ldots, k$.}

\vspace{2.0mm}
\noindent{The skip model reconciles the two extremes: it corresponds to the
linear model for $s=1$ and to the exponential model for $s=k$.
In Section~\ref{sec:use_cases}, we evaluate the time and space complexity of
these models for a set of iterative programs.}

\section{Incremental Processing} \label{sec:incremental_processing}
\label{sec:incremental}
% {\bf Goal. }
In this section we develop techniques for converting linear algebra programs
into functionally equivalent {\em incremental programs} suited for execution on
dynamic datasets.
An incremental program consists of a set of triggers,  one trigger for each
input matrix that might change over time.
Each trigger has a list of update statements that maintain the result for
updates to the associated input matrix.
The total execution cost of an incremental program is the sum of execution costs
of its triggers.
Incremental programs incur lower computational complexity by converting the
expensive operations of non-incremental programs to work with smaller datasets.
Incremental programs combine precomputed results with low-rank
updates to avoid costly operations, like matrix-matrix multiplications or matrix
inversions.
\begin{definition} \label{def:rank} A matrix $\mathcal{M}$ of
dimensions $(n \times n)$ is said to have rank-$k$ if the maximum number of
linearly independent rows or columns in the matrix is $k$. $\mathcal{M}$ is called a
low-rank matrix if $k\ll n$.
\end{definition}

\begin{comment}
Input matrices might change over time.
This section provides an in-depth discussion on 
how to efficiently execute programs in those situations.
To achieve this goal, however, we need to know which input matrices 
are changing, and also the frequencies of these changes. 
In our sample program from Example~\ref{example_intro},
suppose that we want to incrementally update
$F$, $G$, $H$ whenever $A$ or $E$ gets changed. 
It is obvious that if $A$ changes, we need to update all the output
matrices; in contrast, changing $E$ affects only $H$.
Thus, changes of the input matrices might incur different computational
complexities.
The update frequencies for these matrices may dictate 
how such programs should be optimized. 
For that purpose, we assign an update frequency to each input matrix, 
denoted by $f_A$ for matrix $A$.
The update frequency of a static input matrix equals zero.
\end{comment}
\begin{comment}
The overall cost of executing such a program is defined to be a weighted sum of 
all the trigger costs; the weights correspond to the update frequencies of the input matrices.
The ultimate goal of our approach is to transform a given program into 
an incremental program with the least possible execution cost,
building upon the techniques for optimization of linear algebra programs 
presented in Section about transformations.
\end{comment}
\subsection{Delta Derivation}
\label{sec:delta_rules}

The basic step in building incremental programs is the derivation of
{\em delta expressions}
% Here, we show how to construct {\em delta expressions}.
$\Delta_{A}(E)$, which capture how the result of an expression $E$ changes as an
input matrix $A$ is updated by $\Delta{A}$. We consider the update
$\Delta{A}$, called a {\em delta matrix}, to be constant
 and independent of any other matrix.
If we represent $E$ as a function of $A$, then $\Delta_{A}(E) = E(A + \Delta{A})
- E(A)$.
For presentation clarity, we omit the subscript in $\Delta_{A}(E)$
when $A$ is obvious from the context.

Most standard operations of linear algebra are amenable to incremental
processing.
Using the distributive and associative properties of common matrix
operations, we derive the following set of delta rules for updates to $A$. 

\vspace{-2mm}
{\footnotesize
\newcommand{\A}[0]{E_1}
\newcommand{\B}[0]{E_2}
\newcommand{\E}[0]{E}
\begin{align*}
&\Du (\A \mul \B) := \ (\Du \A) \mul \B + \A \mul (\Du \B) + 
  \ (\Du \A) (\Du \B) \\
\end{align*}
\vspace{-8mm}
\begin{align*}
%\qquad \qquad
\Du (\A \pm \B) :=& \ (\Du \A) \pm (\Du \B) \\
%\Du (-\E) := &\ -(\Du \E) \\
\Du(\lambda \mul \E) :=& \ \lambda \mul (\Du \E) \\
\Du(\tr{\E\,}) :=& \ \tr{(\Du \E)} \\
\Du(\inv{\E}) :=& \ \inv{(\E + \Du \E)} - \inv{\E}  \\
\Du(A) :=& \ \Delta{A} \\
\Du(B) :=& \ 0 \qquad (A \neq B) %\qquad  \Du(\lambda) := \ 0  
\end{align*}}
%Here, $\lambda$ is a scalar or a delta matrix, thus $\Du(\Delta{A}) = 0$.

\vspace{-4.0mm}
We observe that the delta rule for matrix inversion references the original expression (twice), which implies that it is more expensive to compute the delta expression than the original expression. 
This claim is true for arbitrary updates to $A$ (e.g., random updates of all
entries, all done at once).
Later on, we discuss a special form of updates that admits efficient incremental maintenance of matrix inversions. 
Note that if $A$ does not appear in $E$, the delta expression for matrix inversion is zero.

\vspace{-1.0mm}
\begin{example} \label{ex:ols}
   \em
   This example shows the derivation process.
%    \todo[inline]{Use the numbers from the table to facilitate the derivation}
   Consider the Ordinary Least Squares method for estimating the unknown
   parameters in a linear regression model. We want to find a statistical estimate of the parameter $\beta^*$ best satisfying $Y=X\beta$.
The solution, written as a linear algebra program, is
$\beta^{*} = (\tr{X} \mul X)^{-1} \mul \tr{X} \mul Y$.
Here, we focus on how to derive the delta expression for $\beta^{*}$ under updates to $X$.
We defer an in-depth cost analysis of the method to Section~\ref{sec:use_cases}.
Let $Z = \tr{X} \mul X$ and $W = \inv{Z}$. Then
\begin{align*}
    \Delta{Z} &= \Delta(\tr{X} \mul X)  \\
      &= (\Delta{(\tr{X})}) \mul X + \tr{X} \mul (\Delta{(X)}) + (\Delta{(\tr{X})}) \mul (\Delta{(X)}) \\
      &= \tr{(\Delta{X})} \mul X + \tr{X} \mul (\Delta{X}) + \tr{(\Delta{X})} \mul (\Delta{X}) \\
  \Delta{W} &= \inv{(Z + \Delta{Z})} - \inv{Z}
\end{align*}
\vspace{-6mm}
\begin{align*}
  \Delta{\beta^{*}} &= 
  (\Delta{W}) \mul \tr{X} \mul Y + W \mul \tr{(\Delta{X})} \mul Y + (\Delta{W}) \mul \tr{(\Delta{X})} \mul Y
  \tag*{\punto} 
\end{align*} 
%For random updates to $X$, computing $\Delta{W}$ involves two matrix inversions, which require $\bigO(n^3)$ operations. 
\end{example}

%For updates of a special form, the incremental evaluation of matrix inverse 

For random matrix updates, incrementally computing a matrix inverse is
prohibitively expensive.
The Sherman-Morrison formula~\cite{Press:2007:NRE:1403886} provides a
numerically cheap way of maintaining the inverse of an invertible matrix for
rank-$1$ updates.
Given a rank-$1$ update $\du \mul \tr{\dv}$, where $\du$ and $\dv$ are column
vectors, if $E$ and $E + \du \mul \tr{\dv}$ are nonsingular, then
%If $E$ is nonsingular and the update $\du \mul \tr{\dv}$, where $\du$ and
%$\dv$ are column vectors, is such that $E + \du \mul \tr{\dv}$ is nonsingular,
% then
\begin{align*}
    \Delta{(\inv{E})} := - \frac{\inv{E} \mul \du \mul \tr{\dv} \mul \inv{E}}{1 + \tr{\dv} \mul \inv{E} \mul \du}
\end{align*}
Note that $\Delta{(\inv{E})}$ is also a rank-$1$ matrix. For instance,
$\Delta{(\inv{E})} = p \mul \tr{q}$, where $p = \lambda \mul \inv{E}\du$ and $q=\tr{(\inv{E})} \mul \dv$ are column vectors, and $\lambda$ is a scalar (observe that the denominator is a scalar too).
Thus, incrementally computing $\inv{E}$ for rank-$1$ updates to $E$ requires
$\bigO(n^2)$ operations; it avoids any matrix-matrix multiplication and
inversion operations.

\begin{example} \label{ex:ols_inc}
  \em 
  We apply the Sherman-Morrison formula to Example~\ref{ex:ols}.   
  We start by considering rank-$1$ updates to $X$.
  Let $\Delta{X} = \du \mul \tr{\dv}$, then
  \begin{align*}
    \Delta{Z} 
    =& \dv \mul \tr{\du} \mul X + \tr{X} \mul \du \mul \tr{\dv} + \dv \mul \tr{\du} \mul \du \mul \tr{\dv} \\
    =& \dv \mul (\tr{\du} \mul X) + (\tr{X} \mul \du + \dv \mul \tr{\du} \mul \du) \mul \tr{\dv} 
  \end{align*}
  The parentheses denote the subexpressions that evaluate to vectors. 
  We observe that each of the monomials is a vector outer product. 
  Thus, we can write $\Delta{Z} = \Delta{Z_1} + \Delta{Z_2}$, where $\Delta{Z_1} = p_1 \mul \tr{q_1}$ and $\Delta{Z_2} = p_2 \mul \tr{q_2}$.  
  Now we can apply the formula on each outer product in turn.
  \begin{align*}
    \Delta_{Z_1}{(W)} =& -\frac{W \mul p_1 \mul \tr{q_1} \mul W}{1 + \tr{q_1} \mul W \mul p_1} \\
    \Delta_{Z_2}{(W)} =& -\frac{(W + \Delta_{Z_1}(W)) \mul p_2 \mul \tr{q_2} \mul (W + \Delta_{Z_1}(W))}{1 + \tr{q_2} \mul (W + \Delta_{Z_1}(W)) \mul p_2} 
  \end{align*}
  
  Finally, $\Delta_{Z}{(W)} = \Delta_{Z_1}{(W)} + \Delta_{Z_2}{(W)}$. 
  The evaluation cost of $\Delta_{Z}{(W)}$ is $\bigO(n^2)$ operations, as discussed above. For comparison, the evaluation cost of $\Delta{W}$ in Example~\ref{ex:ols} is $\bigO(n^\gamma)$ operations.
  \punto
\end{example}

In the above OLS examples, $\Delta{W}$ is a matrix with potentially all nonzero
entries.
If we store these entries in a single delta matrix, we still need to perform
$\bigO(n^\gamma)$-cost matrix multiplications in order to compute
$\Delta{\beta^{*}}$. Next, we propose an alternative way of representing delta
expressions that allows us to stay in the realm of $\bigO(n^2)$ computations.

\begin{comment}  
\begin{example} \label{ex:simple_delta}
Consider a simple program consisting of two statements

\begin{Verbatim}[commandchars=\\\{\}]
X = ABC
Y = XX\textsuperscript{\,T}
\end{Verbatim}
where $A$, $B$, and $C$ are square input matrices 
that might change over time, each with a different update frequency.
For simplicity, we focus only on updates to $B$, denoted by $dB$.
Using some basic derivations, we can obtain an incremental version of the program.
\begin{Verbatim}[commandchars=\\\{\}]
ON UPDATE B(dB) \{
    dX := A(dB)C
    dY := (dX)X\textsuperscript{\,T} + X(dX)\textsuperscript{\,T} + (dX)(dX)\textsuperscript{\,T}
    B  += dB
    X  += dX     
    Y  += dY
\}
\end{Verbatim} 
\end{example}
\end{comment}

\subsection{Delta Representation}
\label{sec:delta_representation}

In this section we discuss how to represent delta expressions 
in a form that is amenable to incremental processing. 
This form also dictates the structure of admissible updates to input matrices.
Incremental processing brings no benefit 
if the whole input matrix changes arbitrarily at once.

Let us consider updates of the smallest granularity -- 
single entry changes of an input matrix. The {\em delta matrix} 
capturing such an update contains exactly one nonzero entry being updated. 
The following example shows that even a minor change,
when propagated na\"{i}vely, can cause incremental processing 
to be more expensive than recomputation.

\begin{example}\label{ex:single_cell_delta}
\em
Consider the program of Example~\ref{ex:simple_delta} for computing the fourth power of a given matrix $A$.
Following the delta rules we write 
$\Delta{B} = (\Delta{A}) \mul A + (A + \Delta{A}) \mul (\Delta{A})$.
Fig.~\ref{fig:example_delta_evaluation} shows the effect of a single entry change in $A$ on $\Delta{B}$.
The single entry change has escalated to a change of one row and one column in $B$.
When we propagate this change to the next statement, $\Delta{C}$ becomes a fully-perturbed delta matrix, that is, all entries might have nonzero values. 

Now suppose we want to evaluate $A^8$, so we extend the program with the statement $D := C \mul C$.
To evaluate $\Delta{D}$, which is expressed similarly as $\Delta{B}$,
we need to perform two matrix-matrix multiplications and two matrix additions.
Clearly, in this case, it is more efficient to recompute $D$ using the new $C$ than to incrementally maintain it with $\Delta{D}$.\punto
\end{example}

\vspace{-0.5mm}
The above example shows that linear algebra programs are, in general,
sensitive to input changes. 
Even a single entry change in the input can cause an avalanche effect of perturbations, quickly escalating to its extreme after executing merely two statements.
%Given that each entry of the matrix might have to be changed, the computational complexity of matrix operations has a lower bound of $\bigO(n^2)$.

We propose a novel approach to deal with escalating updates. 
So far, we have used a single matrix to store the result of a delta expression. 
We observe that such representation is highly redundant as delta matrices typically have low ranks. 
Although a delta matrix might contain all nonzero entries, the number of linearly independent rows or columns is relatively small compared to the matrix size.  
In Example~\ref{ex:single_cell_delta}, $\Delta{B}$ has a rank of at most two.

We maintain a delta matrix in a {\em factored form}, represented as a product of two low-rank matrices.
The factored form enables more efficient evaluation of subsequent delta expressions. 
Due to the associativity and distributivity of matrix multiplication, 
we can base the evaluation strategy for delta expressions solely on matrix-vector products, and thus avoid expensive matrix-matrix multiplications.

To achieve this goal, we also represent updates of input matrices in the
factored form.
In this paper we consider rank-$k$ changes of input matrices as they can capture
many practical update patterns.
For instance, the simplest rank-$1$ updates can express perturbations of one
complete row or column in a matrix, or even changes of the whole matrix when the same
vector is added to every row or column.

In Example~\ref{ex:single_cell_delta}, consider a rank-$1$ update $\Delta{A} =
\duA \tr{\dvA}$, where $\duA$ and $\dvA$ are column vectors, then
$\Delta{B} = \duA \mul (\tr{\dvA} \mul A) + (A \mul \duA) \mul \tr{\dvA} + (\duA \mul
    \tr{\dvA} \mul \duA) \mul \tr{\dvA}$
is a sum of three outer products. The parentheses denote the factored
subexpressions (vectors).
The evaluation order enforced by these parentheses yields only matrix-vector and
vector-vector multiplications. Thus, the evaluation of $\Delta{B}$ requires only
$\bigO(n^2)$ operations.

Instead of representing delta expressions as sums of outer products, 
we maintain them in a more compact vectorized form for performance and presentation reasons.
A sum of $k$ outer products is equivalent to a single product of two matrices of sizes $(n \times k)$ and $(k \times n)$, which are obtained by stacking the corresponding vectors together. 
For instance, 
\begin{equation*}
  \du_1 \mul \tr{\dv_1} + \du_2 \mul \tr{\dv_2} + \du_3 \mul \tr{\dv_3} = 
  \begin{bmatrix}
    \du_1 & \du_2 & \du_3
  \end{bmatrix} \mul 
  \begin{bmatrix}
    \tr{\dv_1} \\
    \tr{\dv_2} \\
    \tr{\dv_3} \\
  \end{bmatrix} =
  P \mul \tr{Q}
\end{equation*}
where $P$ and $Q$ are $(n \times 3)$ block matrices.

To summarize, we maintain a delta expression as a product of two low-rank matrices with dimensions $(n \times k)$ and $(k \times n)$, where $k \ll n$.
This representation allows efficient evaluation of subsequent delta expressions without involving expensive $\bigO(n^\gamma)$ operations; instead, we perform only $\bigO(kn^2)$ operations.
A similar analysis naturally follows for rank-$k$ updates with linearly increasing evaluation costs; the benefit of incremental processing diminishes as $k$ approaches the dominant matrix dimension.

%The factored form also admits more involved forms of input updates, 
%like changes of one row or column in a matrix.
%Our approach is, however, not limited only to rank-$1$ updates. The techniques
% presented here can also be leveraged to rank-$k$ updates with linearly increasing evaluation costs; the benefit of %incremental processing diminishes as $k$ approaches the dominant matrix dimension.

Considering low-rank updates of matrices also opens opportunities to benefit from previous work on incrementalizing complex linear algebra operations. 
We have already discussed the Sherman-Morrison method of incrementally
computing the inverse of a matrix for rank-$1$ updates.
Other work~\cite{deng2010multiple, EPFL-REPORT-161468} investigates 
rank-$1$ updates in different matrix factorizations, like SVD and 
Cholesky decomposition. 
We can further use these new primitives to enrich our language, and, consequently,
support more sophisticated programs.

\subsection{Delta Propagation}
\label{sec:delta_propagation}

When constructing incremental programs we propagate delta expressions from one statement to another. 
For delta expressions with multiple monomials, factored representations include increasingly more outer products.
That raises the cost of evaluating these expressions. 
%In this section we discuss how to limit the effect of such an increase.
In Example~\ref{ex:single_cell_delta}, $\Delta{B}$ consists of three outer products compacted as
\begin{align*}
  \Delta{B} =   
  \begin{bmatrix}
    \duA & (A \mul \duA) & (\duA \mul (\tr{\dvA} \mul \duA))
  \end{bmatrix} \mul
  \begin{bmatrix}
    \tr{\dvA} \mul A \\
    \tr{\dvA} \\
    \tr{\dvA}
  \end{bmatrix} 
  =  \dUB \mul \tr{\dVB}  
\end{align*}
Here, $\dUB$ and $\dVB$ are $(n \times 3)$ block matrices. 
Akin to $\Delta{B}$, $\Delta{C}$ is also a sum of three products
expressed using $B$, $\dUB$, and $\dVB$, and compacted as a product of two $(n \times
9)$ block matrices. Finally, we use $C$ and the factored form of $\Delta{C}$ to
express $\Delta{D}$ as a product of two $(n \times 27)$ block matrices.

Observe that $\dUB$ and $\dVB$ have linearly dependent columns, which suggests
that we could have an even more compact representation of these matrices.
A less redundant form, which reduces the size of $\dUB$ and $\dVB$, guarantees
less work in evaluating subsequent delta expressions.
To alleviate the redundancy in representation, we reduce the number of monomials
in a delta expression by extracting common factors among them.
This syntactic approach does not guarantee the most compact representation of a
delta expression, which is determined by the rank of the delta matrix.
However, computing the exact rank of the delta matrix requires inspection of the
matrix values, which we deem too expensive.
The factored form of $\Delta{B}$ of Example~\ref{ex:single_cell_delta} is
\begin{align*}
  \Delta{B} =   
  \begin{bmatrix}
    \duA & (A \mul \duA + \duA \mul (\tr{\dvA} \mul \duA))
  \end{bmatrix} \mul
  \begin{bmatrix}
    \tr{\dvA} \mul A \\
    \tr{\dvA}
  \end{bmatrix} 
  =  \dUB \mul \tr{\dVB}    
\end{align*}
Here, $\dUB$ and $\dVB$ are $(n \times 2)$ matrices. $\Delta{C}$ is 
a product of two $(n \times 4)$ matrices and $\Delta{D}$ multiplies two $(n \times 8)$ matrices.

\subsection{Putting It All Together}

So far we have discussed how to derive, represent, and propagate delta expressions. 
In this section we put these techniques together into an algorithm that compiles a given program to its incremental version.

Alg.~\ref{algo:compile} shows the algorithm that transforms a program
$\mathcal{P}$ into a set of trigger functions $\mathcal{T}$, each of them
handling updates to one input matrix.
For updates arriving as vector outer products, the matching trigger
incrementally maintains the computation result by evaluating a sequence of
assignment statements ($:=$) and update statements ($\pluseq$).

{
\renewcommand{\algorithmicforall}{\textbf{for each}}

\begin{algorithm}[t!]
  \scriptsize
  \caption{Compile program $\mathcal{P}$ into a set of triggers $\mathcal{T}$}
  \label{algo:compile}
%  {\bf Input}: $\mathcal{P} : [\tuple{A_i, E_i}]$,
%  $\mathcal{I} : \{ \tuple{X_i, m_i, n_i} \}$,
%  $\mathcal{O} : \{ Y_i \}$ \\
%  {\bf Output}: 
%  $\mathcal{T} : \{ \tuple{X_i, [\tuple{A_i, E_i}]} \}$
\begin{algorithmic}[1]
  \Function{Compile}{$\mathcal{P}$, $\mathcal{I}$}%{ : $\mathcal{T}$} 
  \State $\mathcal{T} \gets \emptyset$
  \ForAll{$X \in \mathcal{I}$}  
  \State $\mathcal{D} \gets \mbox{list}(\tuple{X, \du, \dv})$
      \ForAll{$\tuple{A_i,E_i} \in \mathcal{P}$}
         \State $\tuple{P_i,Q_i} \gets \Call{ComputeDelta}{E_i, \mathcal{D}}$         
         \State $\mathcal{D} \gets \mathcal{D}.\mbox{append}(\tuple{A_i, P_i, Q_i})$
      \EndFor  
      \State $\mathcal{T} \gets \mathcal{T} \ \cup \, \Call{BuildTrigger}{X, \mathcal{D}}$      
%      \State $T \gets \Call{BuildTrigger}{X, \mathcal{D}}$
%      \State $\mathcal{T} \gets \mathcal{T} \ \cup \, \Call{Optimize}{T, \mathcal{I}, \mathcal{O}}$
  \EndFor 
  \State \Return{$\mathcal{T}$}
  \EndFunction
\end{algorithmic}
\end{algorithm}
}

The algorithm takes as input two parameters:
(1) a program $\mathcal{P}$ expressed as a list of assignment statements, where
each statement is defined as a tuple $\tuple{A_i, E_i}$ of an expression
$E_i$ and a matrix $A_i$ storing the result, and (2) a set of input matrices $\mathcal{I}$,
% $\mathcal{I} = \{ \tuple{X_i, m_i, n_i} \}$, where $X_i$ is of $(m_i \times
% n_i)$ size.
% (3) a set of output matrices $\mathcal{O} = \{ Y_i \}$.
and outputs a set of trigger functions $\mathcal{T}$.

The \textsc{ComputeDelta} function follows the rules from Section~\ref{sec:delta_rules} to derive the delta for a given expression $E_i$ and an update to $X$. 
The function returns two expressions that together form the delta, $\Delta{A_i} = P_i \mul \tr{Q_i}$.
As discussed in Section~\ref{sec:delta_representation}, $P_i$ and $Q_i$ are block matrices in which each block has its defining expression.

The algorithm maintains a list of the generated delta expressions in $\mathcal{D}$. 
Each entry in $\mathcal{D}$ corresponds to one update statement of the trigger program.
The entries respect the order of statements in the original program.

Note that \textsc{ComputeDelta} takes $\mathcal{D}$ as input.
The list of matrices affected by a change in $X$ -- initially containing only $X$ -- expands throughout the execution of the algorithm.
One expression might reference more than one such matrix, so we have to deal with  multiple matrix updates to derive the correct delta expression.
The delta rules presented in Section~\ref{sec:delta_rules} consider only single matrix updates, but we can easily extend them to handle multiple matrix updates.
Suppose $\mathcal{D} = \{ A, B, \ldots \}$ is a set of the affected matrices that also appear in an expression $E$. Then, 
$\Delta_{\mathcal{D}}(E) := \Delta_{A}(E) + \Delta_{(\mathcal{D} \setminus \{A\})}(E + \Delta_{A}(E))$.
The delta rule considers each matrix update in turn. The order of applying the matrix updates is irrelevant.
\begin{example}
  \em
  Consider the expression $E = A \mul B$ and the updates $\Delta{A}$ and $\Delta{B}$. Then,
  \begin{align*}
    \Delta_{\{A,B\}}{(E)} =& \Delta_{A}(E) + \Delta_{B}(E + \Delta_{A}(E)) \\ 
    =& (\Delta{A}) \mul B + \Delta_{B}(A \mul B + (\Delta{A}) \mul B) \\    
    =& (\Delta{A}) \mul B + A \mul (\Delta{B}) + (\Delta{A}) \mul (\Delta{B})
    \tag*{\punto}
  \end{align*} 
\end{example}

The \textsc{BuildTrigger} function converts the derived deltas $\mathcal{D}$ for
updates to $X$ into a trigger program. % like in Example \ref{ex:simple_delta}.
The function first generates the assignment statements that evaluate $P_i$ and $Q_i$ for each delta expression, and then the update statements for each of the affected matrices. 

\vspace{-1.0mm}
{
\newcommand{\vu}[1]{u\textsubscript{\,#1}}
\newcommand{\vv}[1]{v\textsubscript{\,#1}}
\newcommand{\vvt}[1]{v\textsubscript{\,#1}\textsuperscript{\!T}}
\newcommand{\vU}[1]{U\textsubscript{\,#1}}
\newcommand{\vV}[1]{V\textsubscript{\,#1}}
\newcommand{\vVt}[1]{V\textsubscript{\,#1}\textsuperscript{\!T}}
\renewcommand{\mul}[2]{#1\(\,\)#2}

\begin{example}
   \em
   Consider the program that computes the fourth power of a given matrix A, discussed in Example~\ref{ex:simple_delta} and Example~\ref{ex:single_cell_delta}.
   Algorithm~\ref{algo:compile} compiles the program and produces the following trigger for updates to $A$.
  \begin{Verbatim}[commandchars=\\\{\}]
ON UPDATE A BY (\vu{A},\vv{A}): 
    \vU{B} := [ \vu{A}   (\mul{A}{\vu{A}} + \mul{\vu{A}}{(\mul{\vvt{A}}{\vu{A})}}) ];
    \vV{B} := [ (\mul{A\textsuperscript{\,T}}{\vv{A}})   \vv{A} ];
    \vU{C} := [ \vU{B}   (\mul{B}{\vU{B}} + \mul{\vU{B}}{(\mul{\vVt{B}}{\vU{B})}}) ];
    \vV{C} := [ (\mul{B\textsuperscript{\,T}}{\vV{B}})   \vV{B} ];
    A += \mul{\vu{A}}{\vvt{A}};   B += \mul{\vU{B}}{\vVt{B}};   C += \mul{\vU{C}}{\vVt{C}};
\end{Verbatim} 
Here, $\duA$ and $\dvA$ are column vectors, $U_B$, $V_B$, $U_C$, and $V_C$ are block matrices. 
Each delta, including the input change, is a product of two low-rank matrices. 
%For comparison, the trigger of Example~\ref{ex:simple_delta} uses a single delta matrix. 
\punto
\end{example}
}

\begin{table*}[t]
\footnotesize
\begin{center}
{\renewcommand{\arraystretch}{1.3}    
  \begin{tabular}{@{}lllll@{}}
  \toprule
   \textbf{Model} & \textbf{Matrix Powers} & \textbf{Sums of Matrix Powers} & \textbf{General form: $T_{i+1} = A \mul T_{i} + B$} \\
  \midrule
    Linear & 
$P_i = \left\{
\begin{array}{l}
  A \\  A \mul P_{i-1} 
\end{array} 
\right.$
& 
$S_i = \left\{
\begin{array}{l}
  I \\ A \mul S_{i-1} + I 
\end{array}
\right.$
 & 
$T_i = \left\{
\begin{array}{l}
  A \mul T_0 + B \\ A \mul T_{i-1} + B
\end{array}
\right.$ 
&
$\begin{array}{l@{~}l}
   & \mbox{for }\, i = 1 \\
   & \mbox{for }\, i = 2, 3, \ldots, k
\end{array}$
 \\
 \midrule 
    Exponential & 
$P_{i} = \left\{
\begin{array}{l}
  A \\ P_{i/2} \mul P_{i/2} 
\end{array} 
\right.$ 
& 
$S_{i} = \left\{
\begin{array}{l}
  I \\ P_{i/2} \mul S_{i/2} + S_{i/2} 
\end{array}
\right.$
& 
$T_{i} = \left\{
\begin{array}{l}
  A \mul T_0 + B \\ P_{i/2} \mul T_{i/2} + S_{i/2} \mul B 
\end{array} 
\right.$ 
&
$\begin{array}{l@{~}l}
  & \mbox{for }\, i = 1 \\
  & \mbox{for }\, i = 2, 4, 8, \ldots, k
\end{array}$ \\
 \midrule
    Skip-$s$ & 
$P_{i} = \left\{
\begin{array}{l}
  A \\ P_{i/2} \mul P_{i/2} \\ P_{s} \mul P_{i-s} 
\end{array} 
\right.$    
    & 
$S_{i} = \left\{
\begin{array}{l}
  I  \\  
  P_{i/2} \mul S_{i/2} + S_{i/2} \\
  P_s \mul S_{i-s} + S_s 
\end{array} 
\right.$    
    & 
$T_{i} = \left\{
\begin{array}{l}
  A \mul T_0 + B \\
  P_{i/2} \mul T_{i/2} + S_{i/2} \mul B \\
  P_{s} \mul T_{i-s} + S_{s} \mul B 
\end{array} 
\right.$ 
&
$\begin{array}{l@{~}l@{~}l}
   & \mbox{for } i = 1 \\
   & \mbox{for } i = 2, 4, 8, \ldots, s \\
   & \mbox{for } i = 2s, 3s,\ldots, k
\end{array}$ \\
\bottomrule
\end{tabular}}
\end{center}
\vspace{-3ex}
\caption{The computation of matrix powers, sums of matrix powers, and the general iterative computation expressed as recurrence relations. For simplicity of the presentation, we assume that $\log_{2}k$, $\log_{2}s$, and $\frac{k}{s}$ are integers.}
%\caption{The computation of matrix powers, sums of matrix powers, and the general iterative computation $T_{i+1}=A \mul T_{i} + B$ expressed using the iterative models from Section~\ref{sec:linear_programs}. The rightmost conditions apply to every computation. For simplicity of the presentation, we assume that $\log_{2}k$, $\log_{2}s$, and $\frac{k}{s}$ are integers.}
%\vspace{-5ex}
\label{table:iterative_models}
\end{table*}

\section{Incremental Analytics}
\label{sec:use_cases}

In this section we analyze a set of programs that have wide application across
many domains from the perspective of incremental maintenance.
We study the time and space complexity of both re-evaluation and incremental
evaluation over dynamic datasets.
We show analytically that, in most of these examples, incremental maintenance
exhibits better asymptotic behavior than re-evaluation in terms of execution
time. In other cases, a combination of the two strategies offers the lowest time
complexity.
% We refer the reader to our technical report for a detailed analysis of these
% evaluation strategies.
Note that our incremental techniques are general and apply to a broader range of
linear algebra programs than those presented here.

\subsection{Ordinary Least Squares}
\label{subsec:ols}

Ordinary Least Squares (OLS) is a classical method for fitting a curve to data. 
The method finds a statistical estimate of the parameter $\beta^*$ best satisfying $Y=X\beta$. 
Here, $X  = (m \times n)$ is a set of predictors, and $Y = (m \times p)$ is a set of responses that we wish to model via a function of $X$ with parameters $\beta$.
The best statistical estimate is $\beta^* = (X^T X)^{-1}X^TY$. Data
practitioners often build regression models from incomplete or inaccurate
data to gain preliminary insights about the data or to test their hypotheses.
As new data points arrive or measurements become more accurate, incremental maintenance avoids expensive reconstruction of the whole model, saving time and frustration.

First, consider the cost of incrementally computing the matrix inverse for changes in $X$.
Let $Z = \tr{X} \mul X$, $W = \inv{Z}$, and $\Delta{X} = \du \mul \tr{\dv}$.
As derived in Example~\ref{ex:ols},
\begin{align*}
  \Delta{Z} = 
  \begin{bmatrix}
    \dv & (\tr{X} \mul \du + \dv \mul \tr{u} \mul u)
  \end{bmatrix} \mul 
  \begin{bmatrix}
    \tr{\du} \mul X \\
    \tr{\dv} \\
  \end{bmatrix} =
  \begin{bmatrix}
    p_1 & p_2
  \end{bmatrix} \mul 
  \begin{bmatrix}
    \tr{q_1} \\
    \tr{q_2} \\
  \end{bmatrix} 
\end{align*}
The cost of computing $p_2$ and $q_1$ is $\bigO(mn)$. The vectors $p_1$, $q_1$, $p_2$, and $q_2$ have size $(n \times 1)$.

As shown in Example~\ref{ex:ols_inc}, we could represent the delta expressions
of $W$ as a sum of two outer products, $\Delta_{Z_1}(W) = r_1 \mul \tr{s_1}$ and
$\Delta_{Z_2}(W) = r_2 \mul \tr{s_2}$; for instance, $s_1 = \tr{W} \mul q_1$ and $r_1$ is the remaining subexpression in $\Delta_{Z_1}(W)$.

The computation of $r_1$, $q_1$, $r_2$, and $q_2$ involves only matrix-vector $\bigO(n^2)$ operations.
Then the overall cost of incremental maintenance of $W$ is $\bigO(n^2 + mn)$.
For comparison, re-evaluation of $W$ takes $\bigO(n^\gamma + mn^2)$ operations.

Finally, we compute $\Delta{\beta^{*}}$ for updates $\Delta{X} = \du \mul
\tr{\dv}$ and $\Delta{W} = R \mul \tr{S}$, where $R=\begin{bmatrix} r_1 & r_2
\end{bmatrix}$ and $S = \begin{bmatrix} s_1 & s_2 \end{bmatrix}$ are $(n \times
2)$ block matrices and $\Delta{\beta^*} = R \mul \tr{S} \mul \tr{X} \mul Y + W
\mul \dv \mul \tr{\du} \mul Y + R \mul \tr{S} \mul \dv \mul \tr{\du} \mul Y$.
The optimum evaluation order for this expression depends on the size of $X$ and
$Y$. In general, the cost of incremental maintenance of $\beta^{*}$ is
$\bigO(n^2 + mp + np + mn)$. For comparison, re-evaluation of $\beta^{*}$ takes
$\bigO(mnp + n^2\min{(m,p)})$ operations.

Overall, considering both phases, the incremental maintenance of $\beta{^*}$ for
updates to $X$ has lower computation complexity than re-evaluation. This holds even when $Y$ is of small dimension (e.g., vector); the matrix inversion cost still dominates in the re-evaluation method.
The space complexity of both strategies is $\bigO(n^2)$.

\begin{table*}[t]
\footnotesize
\begin{center}
{\renewcommand{\arraystretch}{1.3}    
  \begin{tabular}{llc@{}c@{}c@{}ccc@{}}
  \toprule
  & \textbf{Model} & 
   \multicolumn{2}{c}{\textbf{(Sums of) Matrix Powers}} & \phantom{ab} &
   \multicolumn{3}{c}{\textbf{General form: $T_{i+1} = A \mul T_{i} + B$}} \\
  \cmidrule{3-4} \cmidrule{6-8}  
  & & Re-evaluation & Incremental & \phantom{ab} & Re-evaluation & Incremental
  & Hybrid \\
  \midrule
  \begin{rotate}{90}\hspace{-0.75cm}Time\end{rotate}
  &  Linear & $n^\gamma k$ & $n^2k^2$ & \phantom{ab} & $pn^2k$ & $(n^2 + pn)k^2$ &
    $pn^2k$ \\
% \midrule 
  &  Exponential & $n^\gamma\log{k}$ & $n^2k$ & \phantom{ab} & $(n^\gamma + pn^2)\log{k}$
    & $(n^2 + pn)k$ & $pn^2\log{k} + n^2k$ \\
% \midrule
  &  Skip-$s$ & $n^\gamma(\log{s} + \frac{k}{s})$ & $n^2\frac{k^2}{s}$ & \phantom{ab} &
    $n^\gamma\log{s} + pn^2(\log{s} + \frac{k}{s})$ & $(n^2 + np)\frac{k^2}{s}$ &  $pn^2(\log{s} + \frac{k}{s}) + n^2s$ \\
\midrule
 \begin{rotate}{90}\hspace{-0.75cm}Space\end{rotate}
 & Linear & $n^2$ & $n^2k$ & \phantom{ab} & $n^2 + np$ & $n^2 + knp$ & $n^2 +
 knp$\\
 & Exponential & $n^2$ & $n^2\log{k}$ & \phantom{ab} & $n^2 + np$ & $(n^2 +
 np)\log{k}$ & $(n^2 + np)\log{k}$ \\
 & Skip-$s$ & $n^2$ & $n^2(\log{s} + \frac{k}{s})$ & \phantom{ab} & $n^2 + np$ &
 $(n^2 + np)\log{s} + np\frac{k}{s}$ & $(n^2 + np)\log{s} + np\frac{k}{s}$ \\
\bottomrule
\end{tabular}}
\end{center}
\vspace{-3ex}
\caption{The time and space complexity (expressed in big-O notation) of the
different evaluation techniques for the various computational models under
rank-$1$ updates to matrix $A$ where $2\leq\gamma\leq3$ as described in Section~\ref{sec:linear_programs}. }
%\caption{The time and space complexity in big-O notation of the re-evaluation,
% incremental, and hybrid evaluation strategies for the computation of matrix
% powers $A^k$, sums of matrix powers, and the general iterative computation $T_{i+1} = A \mul T_{i} + B$, expressed using different iterative models, for rank-$1$ updates to $A$.}
%\vspace{-5ex}
\label{table:time_space_cost}
\end{table*}

\subsection{Matrix Powers}
\label{subsec:matrix_powers}
Our next analysis includes the computation of $A^k$ of a square matrix $A$ for
some fixed $k > 0$.
Matrix powers play an important role in many different domains including
computing the stochastic matrix of a Markov chain after $k$ steps, solving
systems of linear differential equations using matrix exponentials, answering
graph reachability queries where $k$ represents the maximum path length, and
computing PageRank using the power method.

Matrix powers also provide the basis for the incremental analysis of programs having more general forms of iterative computation.
In such programs we often decide to evaluate several iteration steps at once for performance reasons, and matrix powers allow us to express these compound transformations between iterations, as shown later on.

\subsubsection{Iterative Models}

Table~\ref{table:iterative_models} expresses the matrix power computation using the iterative models presented in Section~\ref{sec:linear_programs}.
In all cases, $A$ is an input matrix that changes over time, and $P_k$ contains the final result $A^k$. 
The linear model computes the result of every iteration, while the exponential model makes progressively larger leaps between consecutive iterations evaluating only $\log_{2}{k}$ results.
The skip model precomputes $A^s$ in $P_s$ using the exponential model and  then
reuses $P_s$ to compute every $s^{th}$ subsequent iteration.

\begin{comment}
\noindent
{\bf Linear Model.}
\begin{align*}
P_i = \left\{
\begin{array}{l@{~}l@{~}l}
  A & \quad & \mbox{if }\, i = 1 \\
  A \mul P_{i-1} & \quad & \mbox{if }\, i = 2, 3, \ldots k
\end{array}
\right.%, \mbox{ for } i = 1, 2, \ldots k
\end{align*}

\noindent
{\bf Exponential Model.}
\begin{align*}  
P_{i} = \left\{
\begin{array}{l@{~}l@{~}l}
  A & \quad & \mbox{if }\, i = 1 \\
  P_{i/2} \mul P_{i/2} & \quad & \mbox{if }\, i = 2, 4, \ldots, k
\end{array}
\right.%, \mbox{ for } i = 1, 2, 4,  \ldots k
\end{align*}

\noindent
{\bf Skip Model.}
\begin{align*}  
P_{i} = 
\left\{
\begin{array}{l@{~}l@{~}l}
  A & \quad & \mbox{if }\, i = 1 \\
  P_{i/2} \mul P_{i/2} & \quad & \mbox{if }\, i = 2, 4, \ldots, s \\
  P_{s} \mul P_{i-s} & \quad & \mbox{if }\, i = 2s, 3s, \ldots, k
\end{array}
\right.%, \mbox{ for } i = 1, 2, 4, \ldots s, 2s, 3s, \ldots k
\end{align*}
\end{comment}

Expressing the matrix power computation as an iterative process eases the complexity analysis of both re-evaluation and incremental maintenance, which we show next.

\subsubsection{Cost Analysis}

We analyze the time and space complexity of re-evaluation and incremental
maintenance of $P_k$ for rank-$1$ updates to $A$, denoted by $\Delta{A} = \du
\mul \tr{\dv}$.
We assume that $A$ is a dense square matrix of size $(n \times n)$. 
% TODO: special cases of rank-$1$ updates

\noindent{\bf Re-evaluation.}\ Table~\ref{table:time_space_cost} shows the time complexity of re-evaluating
$P_k$ in different iterative models.
The re-evaluation strategy first updates $A$ by $\Delta{A}$ and then recomputes $P_k$ using the new value of $A$. 
All three models perform one $\bigO(n^\gamma)$ matrix-matrix multiplication per iteration. 
The total execution cost thus depends on the number of iteration steps:  
The exponential method clearly requires the fewest iterations $\log_{2}{k}$, followed by $(\log_{2}s + \frac{k}{s})$ and $k$ iterations of the skip and linear models.

\begin{comment}
\begin{table}[t]
\begin{center}
{\renewcommand{\arraystretch}{1.3}  
\begin{tabular}{l | c c c }

%  \begin{rotate}{90}\hspace{-1.2cm}Time \end{rotate} 
  &  Linear & Exponential & Skip  \\ 
  \hline
  Re-evaluation & $n^\gammak$ & $n^\gamma\log{k}$ & $n^\gamma(\log{s} + \frac{k}{s})$ \\
  Incremental & $n^2k^2$ & $n^2k$ & $n^2\frac{k^2}{s}$ \\  
  \hline
\end{tabular}}
\vspace{-1ex}
\caption{The time complexity in big-O notation of re-evaluation and incremental maintenance  
of the iterative models for computing $A^k$ after a rank-$1$ update to $A$. 
}
\label{table:matrix_power_exec_cost}
\end{center}
\end{table}
\end{comment}

Table~\ref{table:time_space_cost} also shows the space complexity of
re-evaluation in the three iterative models.
The memory consumption of these models is independent of the number of iterations. At each iteration step these models use at most two previously computed values, but not the full history of $P_i$ values.
%, plus the input matrix $A$. 

\begin{comment}
\begin{table}[t]
\begin{center}
\vspace{-4.5ex}
{\renewcommand{\arraystretch}{1.3} 
\begin{tabular}{ l | c c c }
  ~ &  Linear & Exponential & Skip   \\
  \hline
%  Reevaluation & $3n^2$ & $4n^2$ & $3n^2$ \\
  Re-evaluation & $n^2$ & $n^2$ & $n^2$ \\
  Incremental & $n^2k$ & $n^2\log{k}$ & $n^2(\log{s} + \frac{k}{s})$ \\
%  Incremental & $kn^2$ & $(\lceil\log_{2}s\rceil + \left\lceil\frac{k}{s}\right\rceil)n^2$ & $(\lceil\log_{2}k\rceil + 1)n^2$ \\
  \hline
\end{tabular}}
\vspace{-1ex}
\caption{The space complexity in big-O notation of re-evaluation and incremental
maintenance of the iterative models for computing $A^k$ after a rank-$1$ update
of $A$.
%Assumption: $\mbox{log}_{2}k$, $\mbox{log}_{2}s$, and $\frac{k}{s}$ are integers.
}
\vspace{-5ex}
\label{table:matrix_power_mem_cost}
\end{center}
\end{table}
\end{comment}

\noindent{\bf Incremental Maintenance.}\ This strategy captures the change in
the result of every iteration as a product of two low-rank matrices $\Delta{P_i} = \dU_i \mul \tr{\dV_i}$.
The size of $\dU_i$ and $\dV_i$ and, in general, the rank of $\Delta{P_i}$ grow linearly with every iteration step.
We consider the case when $k \ll n$ in which we can profit from the low-rank delta representation. 
This is a realistic assumption as many practical computations consider large
matrices and relatively few iterations; for example, 80.7\% of the pages in a 
PageRank computation converge in less than 15
iterations~\cite{Kamvar03adaptivemethods}.

%Incremental computation of matrix powers proceeds in two phases. 
%The first phase computes $U_i$ and $V_i$ for every iteration $i$, based on the old results and deltas of previous iterations. 
%The second phase updates $T_i$ by the product of $U_i$ and $V_i$.
%Thus, one iteration consists of three different computations: $U_i$, $V_i$, and $U_i \mul \tr{V_i}$.

Table~\ref{table:time_space_cost} shows the time complexity of incremental maintenance of $P_k$ in the three iterative models.
Incremental maintenance exhibits better asymptotic behavior than re-evaluation in all three models, with the exponential model clearly dominating the others. 
Appendix~\ref{sec:appendix_matrix_powers} presents a detailed cost analysis.

The performance improvement comes at the cost of increased memory consumption, as incremental maintenance requires storing the result of every iteration step.
Table~\ref{table:time_space_cost} also shows the space complexity of incremental
maintenance for the three iterative models.
%Overall, incremental maintenance achieves better asymptotic behavior than naive reevaluation but at the price of increased memory complexity. 

\subsubsection{Sums of Matrix Powers}
\label{subsec:sums_matrix_powers}

A form of matrix powers that frequently occurs in iterative computations is a sum of matrix powers. 
The goal is to compute $S_k = I + A + \ldots A^{k-2} + A^{k-1}$, for a given matrix $A$ and fixed $k > 0$. Here, $I$ is the identity matrix.

In Table~\ref{table:iterative_models} we express this computation using the
iterative models discussed earlier. In the exponential and skip models, the computation of $S_k$ relies on the results of matrix power computation, denoted by $P_i$ and evaluated using the exponential model discussed earlier.

\begin{comment}
\noindent
{\bf Linear Model.} 
\begin{align*}
S_i &= \left\{
\begin{array}{l@{~}l@{~}l}
  I & \quad & \mbox{if }\, i = 1 \\
  S_{i-1} \mul A + I & \quad & \mbox{if }\, i = 2, 3, \ldots, k
\end{array}
\right.%, \mbox{ for } i = 1, 2, \ldots k
\end{align*}

\noindent
{\bf Exponential Model.}
\begin{align*}  
S_{i} &= \left\{
\begin{array}{l@{~}l@{~}l}
  I & \quad & \mbox{if }\, i = 1 \\
  P_{i/2} \mul S_{i/2} + S_{i/2} & \quad & \mbox{if }\, i = 2, 4, \ldots, k
\end{array}
\right.%, \mbox{ for } i = 1, 2, 4,  \ldots k 
\end{align*}

\noindent
{\bf Skip Model.}
\begin{align*}  
S_{i} &= \left\{
\begin{array}{l@{~}l@{~}l}
  I & \quad & \mbox{if }\, i = 1 \\  
  P_{i/2} \mul S_{i/2} + S_{i/2} & \quad & \mbox{if }\, i = 2, 4, \ldots, s \\
  P_s \mul S_{i-s} + S_s & \quad & \mbox{if }\, i = 2s, 3s, \ldots, k
\end{array}
\right.%, \mbox{ for } i = 1, 2, 4,  \ldots k 
\end{align*}
\end{comment}

For all three models, the time and space complexity of computing sums of matrix powers is the same in terms of big-O notation as that of computing matrix powers.
The intuition behind this result is that the complexity of each iteration step has remained unchanged. Each iteration step performs one matrix addition more, but the execution cost is still dominated by the matrix multiplication.
We omit the detailed analysis due to space constraints.

\subsection{General Form: T\textsubscript{\bf{i+1}} = AT\textsubscript{\bf{i}} + B}
\label{subsec:general_form}

The two examples of matrix power computation provide the basis for the
discussion about a more general form of iterative computation: $T_{i+1} = A \mul
T_{i} + B$, where $A$ and $B$ are input matrices.
In contrast to the previous analysis of matrix powers, this iterative
computation involves also non-square matrices, $T = (n \times p)$, $A = (n
\times n)$, and $B=(n \times p)$, making the choice of the optimum evaluation
strategy dependent on the values of $n$, $p$, $k$, and $s$.

Many iterative algorithms share this form of computation including gradient
descent, PageRank, iterative methods for solving systems of linear
equations, and the power iteration method for eigenvalue computation.
Here, we analyze the complexity of the general form of iterative computation,
and the same conclusions hold in all these cases.

\subsubsection{Iterative Models}

The iterative models of the form $T_{i+1} = A \mul T_{i} + B$, which
are presented in Table~\ref{table:iterative_models}, rely on the computations of
matrix powers and sums of matrix powers.
To understand the relationship between these computations, consider the iterative process $T_{i+1} = A \mul T_{i} + B$ that has been ``unrolled'' for $k$ iteration steps. The direct formula for computing $T_{i+k}$ from $T_{i}$ is
$T_{i+k} = A^k \mul T_i + (A^{k-1} + \ldots + A + I) \mul B$.

We observe that $A^k$ and $\sum_{i=0}^{k-1}{A^i}$ correspond to $P_k$ and $S_k$ in the earlier examples, for which we have already shown efficient (incremental) evaluation strategies.
Thus, to compute $T_i$, we maintain two auxiliary views $P_i$ and $S_i$ evaluating matrix powers and sums of matrix powers using the exponential model discussed before.

\begin{comment}
\noindent
{\bf Linear Model.}
\begin{align*}
T_i = \left\{
\begin{array}{l@{~}l@{~}l}
  A \mul T_0 + B & \quad & \mbox{if }\, i = 1 \\
  A \mul T_{i-1} + B & \quad & \mbox{if }\, i = 2, 3, \ldots, k
\end{array}
\right.%, \mbox{ for } i = 1, 2, \ldots k
\end{align*}

\noindent
{\bf Exponential Model.}
\begin{align*}  
T_{i} =& \left\{
\begin{array}{l@{~}l@{~}l}
  A \mul T_0 + B& \quad & \mbox{if }\, i = 1 \\
  P_{i/2} \mul T_{i/2} + S_{i/2} \mul B & \quad & \mbox{if }\, i = 2, 4, \ldots, k
\end{array}
\right.%, \mbox{ for } i = 1, 2, 4,  \ldots k
\end{align*}

\noindent
{\bf Skip Model.}
\begin{align*}  
T_{i} = \left\{
\begin{array}{l@{~}l@{~}l}
  A \mul T_0 + B & \quad & \mbox{if }\, i = 1 \\
  P_{i/2} \mul T_{i/2} + S_{i/2} \mul B & \quad &  \mbox{if }\, i = 2, 4, \ldots s \\
  P_{s} \mul T_{i-s} + S_{s} \mul B & \quad & \mbox{if }\, i = 2s, 3s, \ldots, k
\end{array}
\right.%, \mbox{ for } i = 1, 2, 4, \ldots s, 2s, \ldots k
\end{align*}
\end{comment}

\subsubsection{Cost Analysis}\label{subsubsec:general_form_cost_analysis}

We analyze the time and space complexity of re-evaluation and incremental
maintenance of $T_k$ for rank-$1$ updates to $A$, denoted by $\Delta{A} = \du
\mul \tr{\dv}$.
We assume that $A$ is an $(n \times n)$ dense matrix and that $T_i$ and $B$ are $(n \times p)$ matrices.
We omit a similar analysis for changes in $B$ due to space constraints.

We also analyze a combination of the two strategies, called hybrid evaluation, which avoids the factorization of delta expressions but instead represents them as single matrices.
We consider this strategy because the size $(n \times p)$ of the delta matrix
$\Delta{T_i}$ might be insufficient to justify the use of the factored form.
For instance, consider an extreme case when $T_i$ is a column vector ($p=1$), then $\Delta{T_i}$ has rank $1$, and further decomposition into a product of two matrices would just increase the evaluation cost.
In such cases, hybrid evaluation expresses $\Delta{T_i}$ as a single matrix and propagates it to the subsequent iterations.

Table~\ref{table:time_space_cost} presents the time complexity of re-evaluation,
incremental and hybrid evaluation of the $T_{i+1} = A \mul T_{i} + B$
computation expressed in different iterative models for rank-$1$ updates to $A$.
The same complexity results hold for the special form of iterative computation where $B = 0$.
We discuss the results for each evaluation strategy next.
Appendix~\ref{sec:appendix_atb} shows a detailed cost analysis.

\begin{comment}
\begin{table}[t]
\begin{center}
{\renewcommand{\arraystretch}{1.3}  
\setlength{\tabcolsep}{4pt}
\vspace{-3.5ex}
\begin{tabular}{c | c c c } 
%  \hline
%   & \multicolumn{3}{c}{Time complexity}\\
%  \begin{rotate}{90}\hspace{-1.2cm}Time \end{rotate} 
  &  Linear & Exponential & Skip  \\
  \hline
  \hspace{-1.5mm}R & $pn^2k$ & $(n^\gamma + pn^2)\log{k}$ & 
  $n^\gamma\log{s} + pn^2(\log{s} + \frac{k}{s})$\\   
  \hspace{-1.5mm}I & $(n^2 + pn)k^2$ & $(n^2 + pn)k$ & $(n^2 + np)\frac{k^2}{s}$\\  
  \hspace{-1.5mm}H & $pn^2k$ & $pn^2\log{k} + n^2k$ & $pn^2(\log{s} + \frac{k}{s}) + n^2s$\\   
  \hline  
\end{tabular}}
\vspace{-1ex}
\caption{The time complexity in big-O notation of re-evaluation, incremental, and hybrid evaluation  
of the iterative models for computing $T=A \mul T + B$ after a rank-$1$ update
to $A$.
}
\vspace{-4ex}
\label{table:atb_exec_cost}
\end{center}
\end{table}
\end{comment}

Table~\ref{table:time_space_cost} also shows the space complexity of the three
iterative models when executed using different evaluation strategies.
The re-evaluation strategy maintains the result of $T_i$, and if needed $P_i$ 
and $S_i$, only for the current iteration; it also stores the input
matrices $A$ and $B$.
%Thus, the memory consumption is independent of the number of iterations. 
In contrast, the incremental and hybrid evaluation strategies materialize the
result of every iteration, thus the memory consumption depends on the number of performed iterations.

\begin{comment}
\begin{table}[t]
\begin{center}
{\renewcommand{\arraystretch}{1.3}   
\setlength{\tabcolsep}{4pt}
\begin{tabular}{ l | c c c }
  \noalign{\smallskip}
  ~ &  Linear & Exponential & Skip   \\
  \hline  
  R & $n^2 + np$ & $n^2 + np$ & $n^2 + np$ \\
  I \& H & $n^2 + knp$ & $(n^2 + np)\log{k}$ & $(n^2 + np)\log{s} + \frac{k}{s}np$ \\
  \hline
\end{tabular}}
\caption{The space complexity in big-O notation of re-evaluation, incremental,
and hybrid evaluation of the iterative models for computing $T=A \mul T + B$
after a rank-$1$ update of $A$.
%Assumption: $\mbox{log}_{2}k$, $\mbox{log}_{2}s$, and $\frac{k}{s}$ are integers.
}
\label{table:atb_mem_cost}
\end{center}
\end{table}
\end{comment}

\noindent{\bf Re-evaluation.}\ The choice of the iterative model with the best asymptotic behavior depends on the value of parameters $n$, $p$, $k$, and $s$. 
The time complexities from Table~\ref{table:time_space_cost} shows that the
linear model incurs the lowest time complexity when $p \ll n$, otherwise the exponential model dominates the others in terms of the running time.
We analyze the cost of each iterative model next.
Note that the re-evaluation strategy first updates $A$ by $\Delta{A}$ and then recomputes $T_i$, and if needed $P_i$ and $S_i$, using the new value of $A$.

\vspace{-0.5mm}
\smartparagraph{$\bullet$ Linear model.}{ The computation performs $k$ iterations, where each one incurs the cost of $\bigO(pn^2)$, and thus the total cost is $\bigO(pn^2k)$.}

%\vspace{-0.5mm}
\smartparagraph{$\bullet$ Exponential model.}{ Maintaining $P_i$ and $S_i$ takes $\bigO(n^\gamma)$ operations as discussed before, while recomputing $T_i$ requires $\bigO(pn^2)$ operations. Overall, the re-evaluation cost is $\bigO((n^\gamma + pn^2)log{k})$.}

%\vspace{-0.5mm}
\smartparagraph{$\bullet$ Skip model.}{ The skip model combines the above two  
models. 
%which reflects on the cost analysis. 
Maintaining $P_s$ and $S_s$ takes $\bigO(n^\gamma\log{s})$ operations as shown
earlier, while recomputing $T_i$ costs $\bigO(pn^2)$ per iteration. The total
number of $(\log_{2}{s} + \frac{k}{s})$ iterations yields the total cost of
$\bigO(n^\gamma\log{s} + pn^2(\log{s} + \frac{k}{s}))$.}

\vspace{0.5mm}
\noindent{\bf Incremental Maintenance.}\ Table~\ref{table:time_space_cost} shows
that incremental evaluation of $T_k$ using the exponential model incurs the
lowest time complexity among the three iterative models.
%(see Appendix~\ref{sec:appendix_atb} for a detailed analysis).
It also outperforms complete re-evaluation when $p > n$, but 
the performance benefit diminishes as $p$ becomes smaller than $n$.
For the extreme case when $p = 1$, 
complete re-evaluation and incremental maintenance have the same asymptotic
behavior, but in practice re-evaluation performs fewer operations as it avoids
the overhead of computing and propagating the factored deltas.
We show next how to combine the best of both worlds to lower the execution time when $p \ll n$.

\vspace{0.5mm} 
\noindent{\bf Hybrid evaluation.}\ Hybrid evaluation departs from incremental maintenance in that it represents the change in the result of every iteration as a single matrix instead of an outer product of two vectors. 
The benefit of hybrid evaluation arises when the rank of $\Delta{T_i} = (n
\times p)$ is not large enough to justify the use of the factored form; that is,
when the dimension $p$ or $n$ is comparable with $k$.

Table~\ref{table:time_space_cost} shows the time complexity of hybrid evaluation
of the $T_{i+1} = A \mul T_{i} + B$ computation expressed in different iterative
models for rank-$1$ updates to $A$. 
For the extreme case when $p=1$, the skip model performs $\bigO(\log{s} + \frac{k}{s} + s)$ matrix-vector multiplications.
In comparison, re-evaluation and incremental maintenance perform $\bigO(k)$ such operations. 
Thus, the skip model of hybrid evaluation bears the promise of better performance for the given values of $k$ and $s$.
%We experimentally evaluate this conjecture in our technical report. %Section~\ref{sec:experiments}.
  
\section{System Overview}
\label{sec:systems}

\begin{figure}[t]
\centering
\includegraphics[width=1.0\columnwidth]{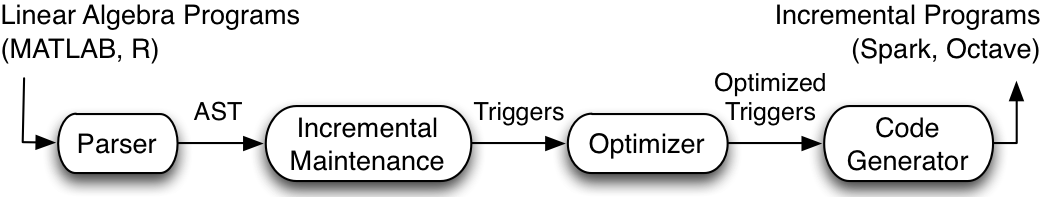}
\vspace{-5mm}
\caption{The \system system overview}
\label{fig:system_overview}
\vspace{-3mm}
\end{figure}

\begin{figure*}[t!]
        \centering
        \begin{subfigure}[b]{0.245\textwidth}
                \centering                
                \includegraphics[width=\textwidth]{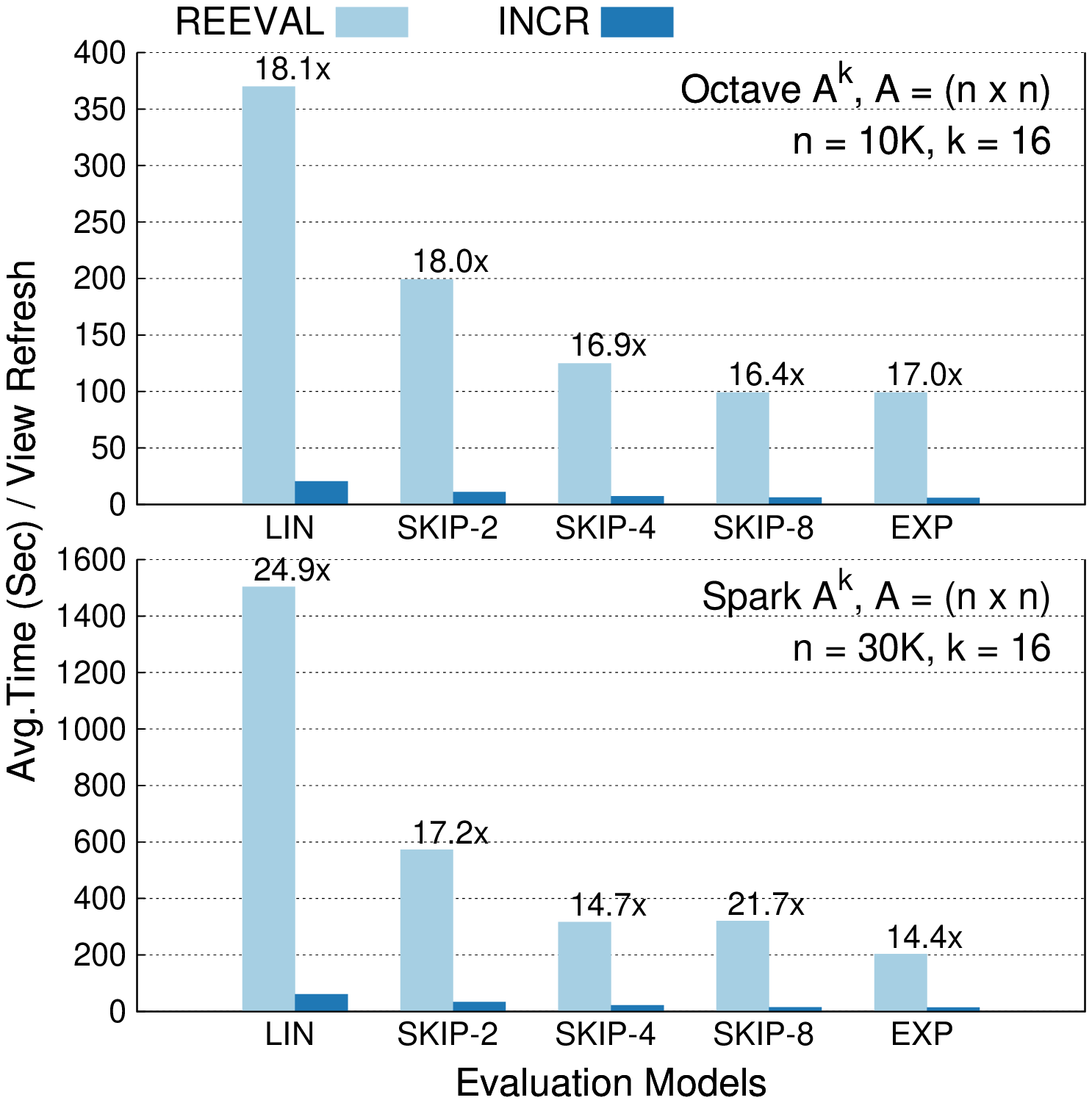}
                \vspace{-5mm}
                \caption{Matrix Powers}
		            \label{fig:mp1}
		            \vspace{1.5mm}
        \end{subfigure}
        \begin{subfigure}[b]{0.245\textwidth}
                \centering
                \includegraphics[width=\textwidth]{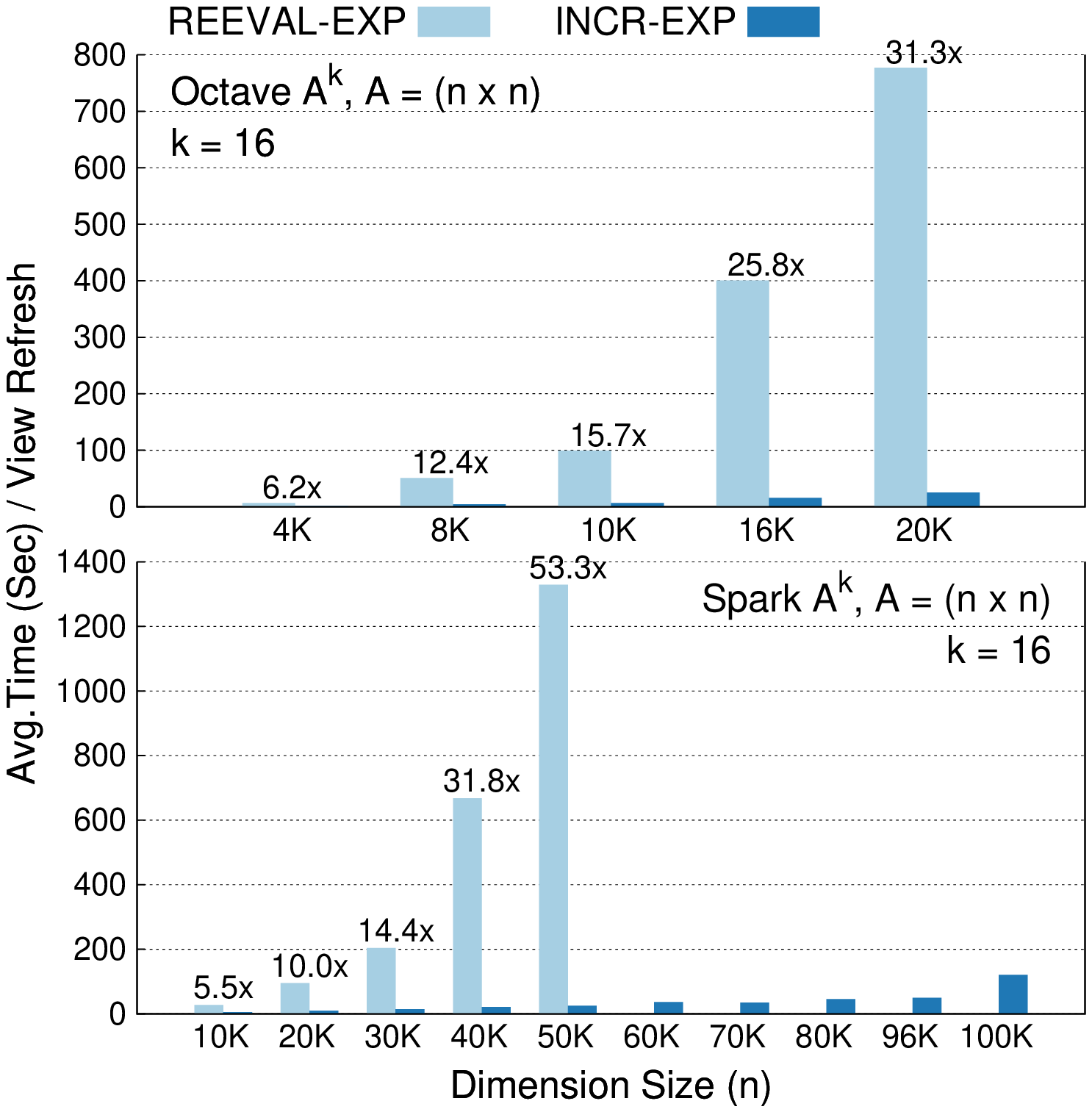}
                \vspace{-5mm}
                \caption{Scalability of Powers $(n)$}
		            \label{fig:mp2}
		            \vspace{1.5mm}
        \end{subfigure}
        \begin{subfigure}[b]{0.245\textwidth}
                \centering
                \includegraphics[width=\textwidth]{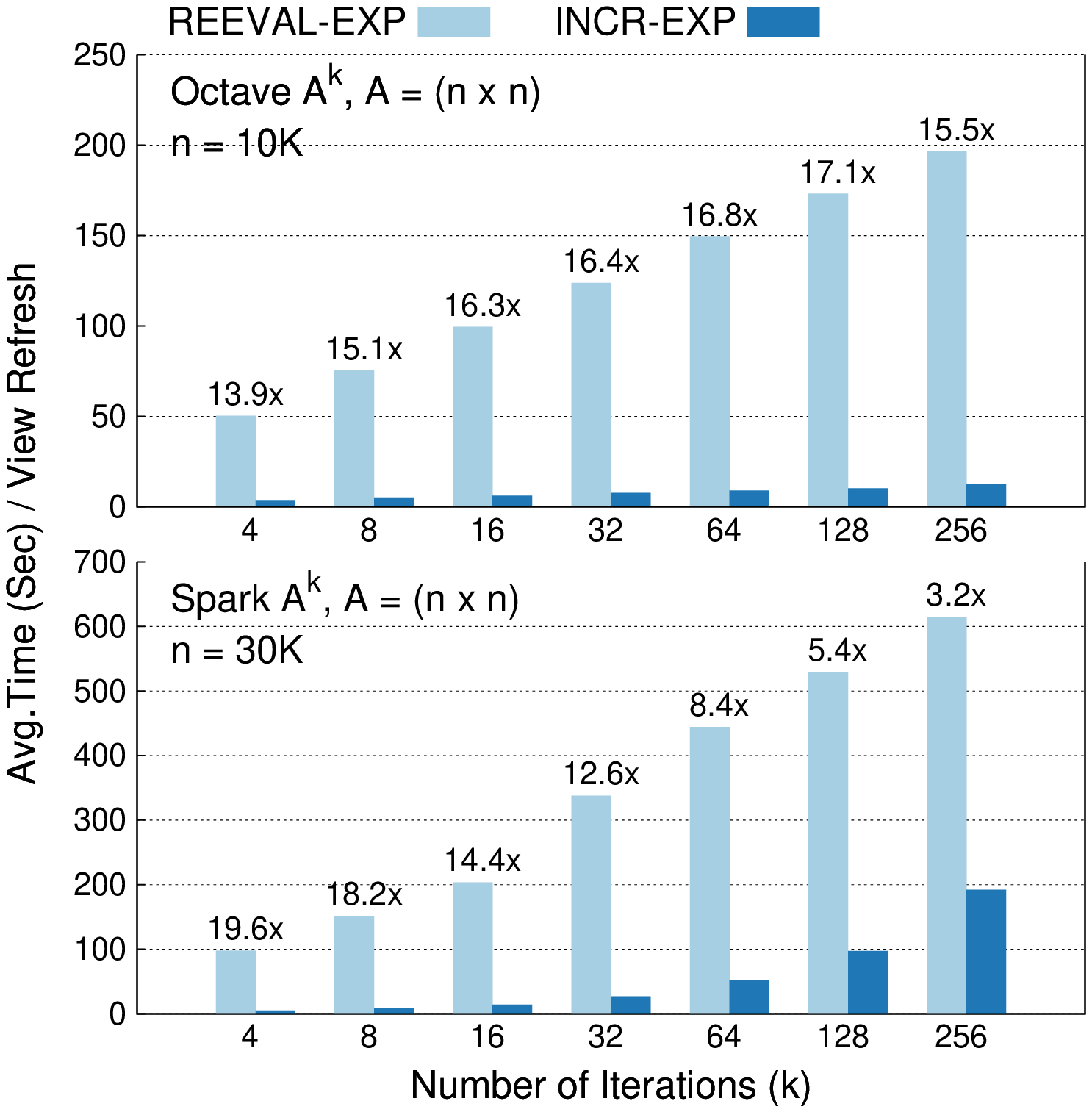}
                \vspace{-5mm}
                \caption{Scalability of Powers $(k)$}
		            \label{fig:mp3}
		            \vspace{1.5mm}
        \end{subfigure}
	      \centering
        \begin{subfigure}[b]{0.245\textwidth}
		            \centering
	              \includegraphics[width=\textwidth]{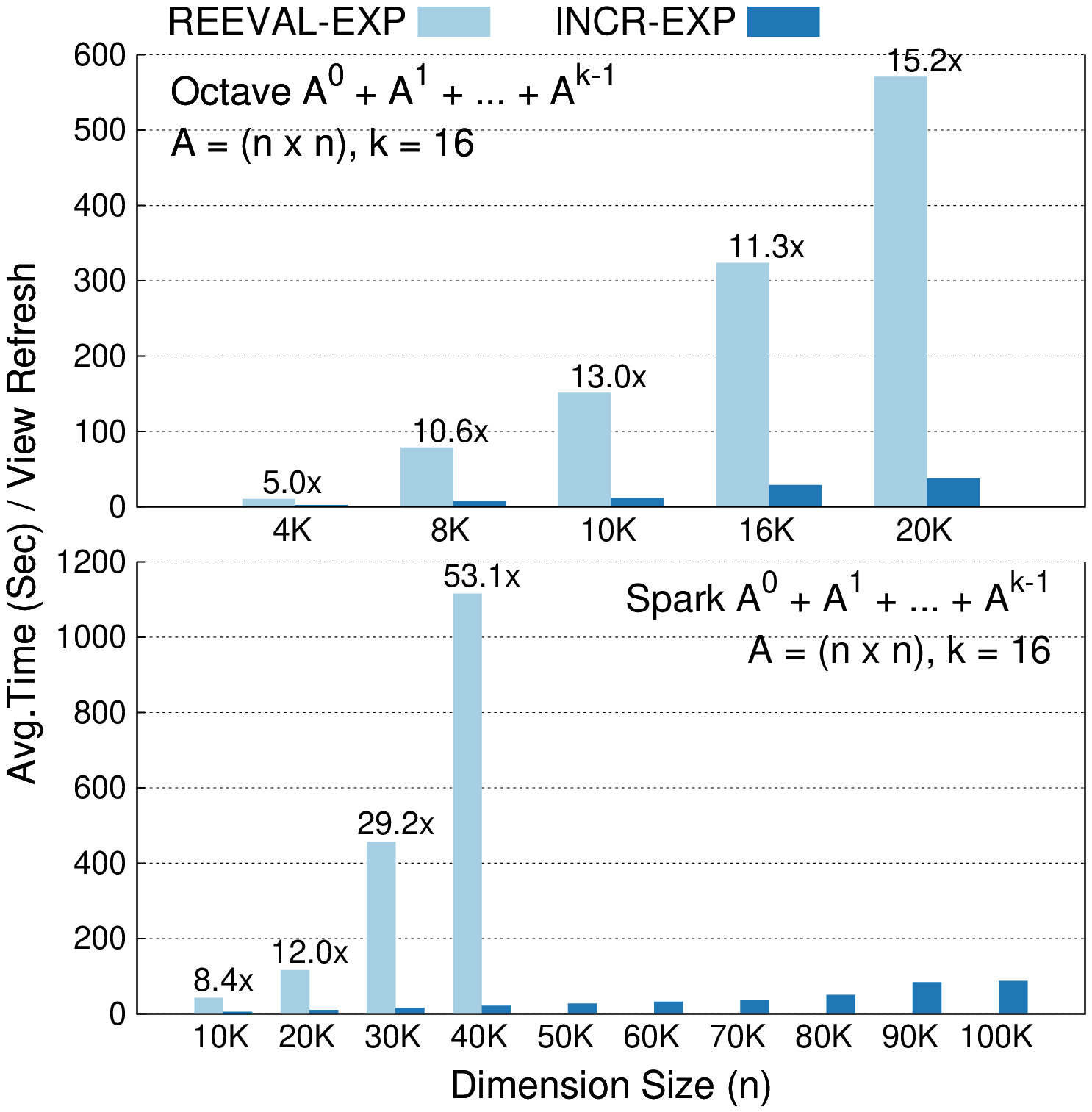}
	              \vspace{-5mm}
	              \caption{Sums of Powers}
		            \label{fig:Sums}
		            \vspace{1.5mm}
        \end{subfigure}
 	      \begin{subfigure}[b]{0.245\textwidth}
                \centering
                \includegraphics[width=\textwidth]{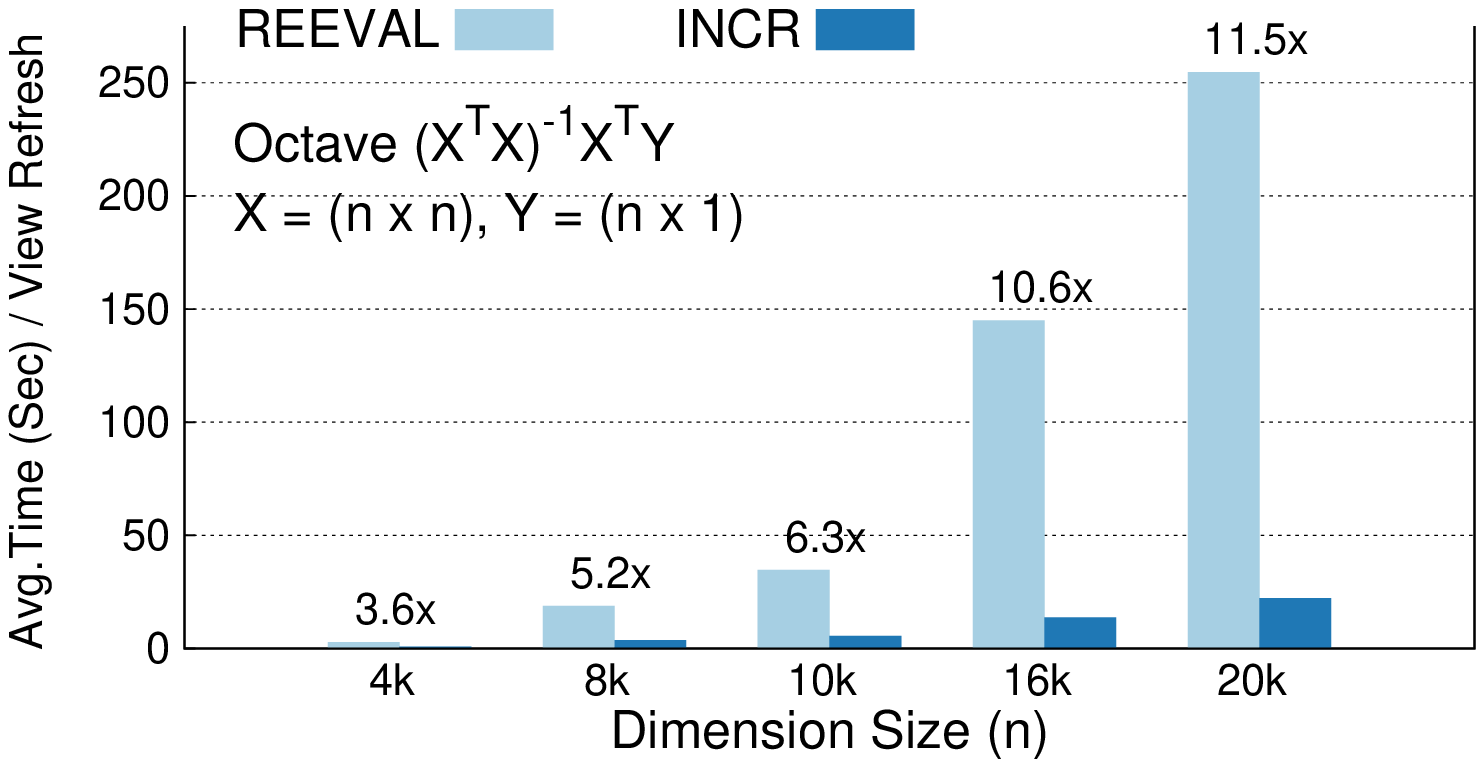}
                \vspace{-5mm}
                \caption{Ordinary Least Squares}
		            \label{fig:ols}
        \end{subfigure}
        \begin{subfigure}[b]{0.245\textwidth}
                \centering
                \includegraphics[width=\textwidth]{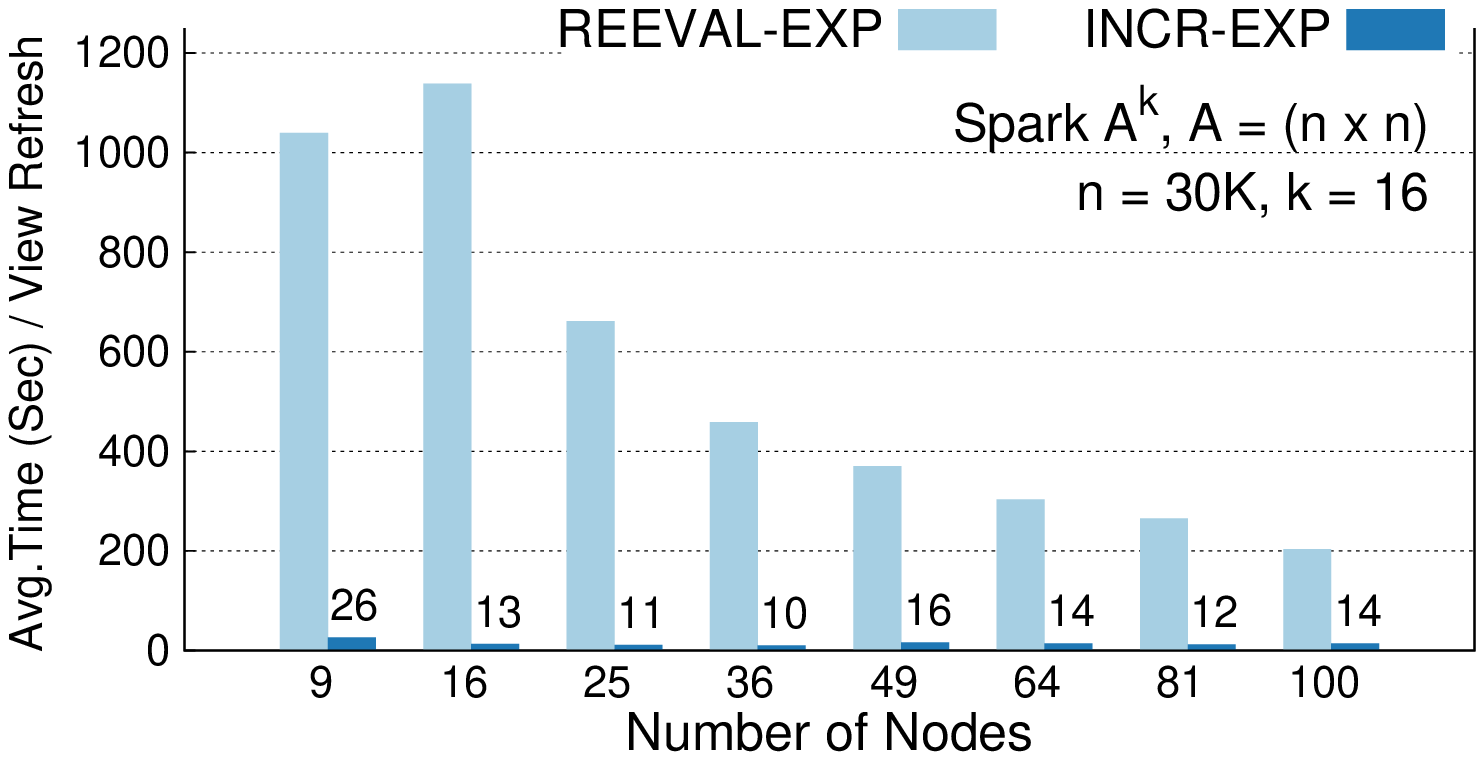}
                \vspace{-5mm}
                \caption{Scalability of Powers (nodes)}
		            \label{fig:mp_scalability_nodes}
       \end{subfigure}
       \begin{subfigure}[b]{0.245\textwidth}
                \centering
                \includegraphics[width=\textwidth]{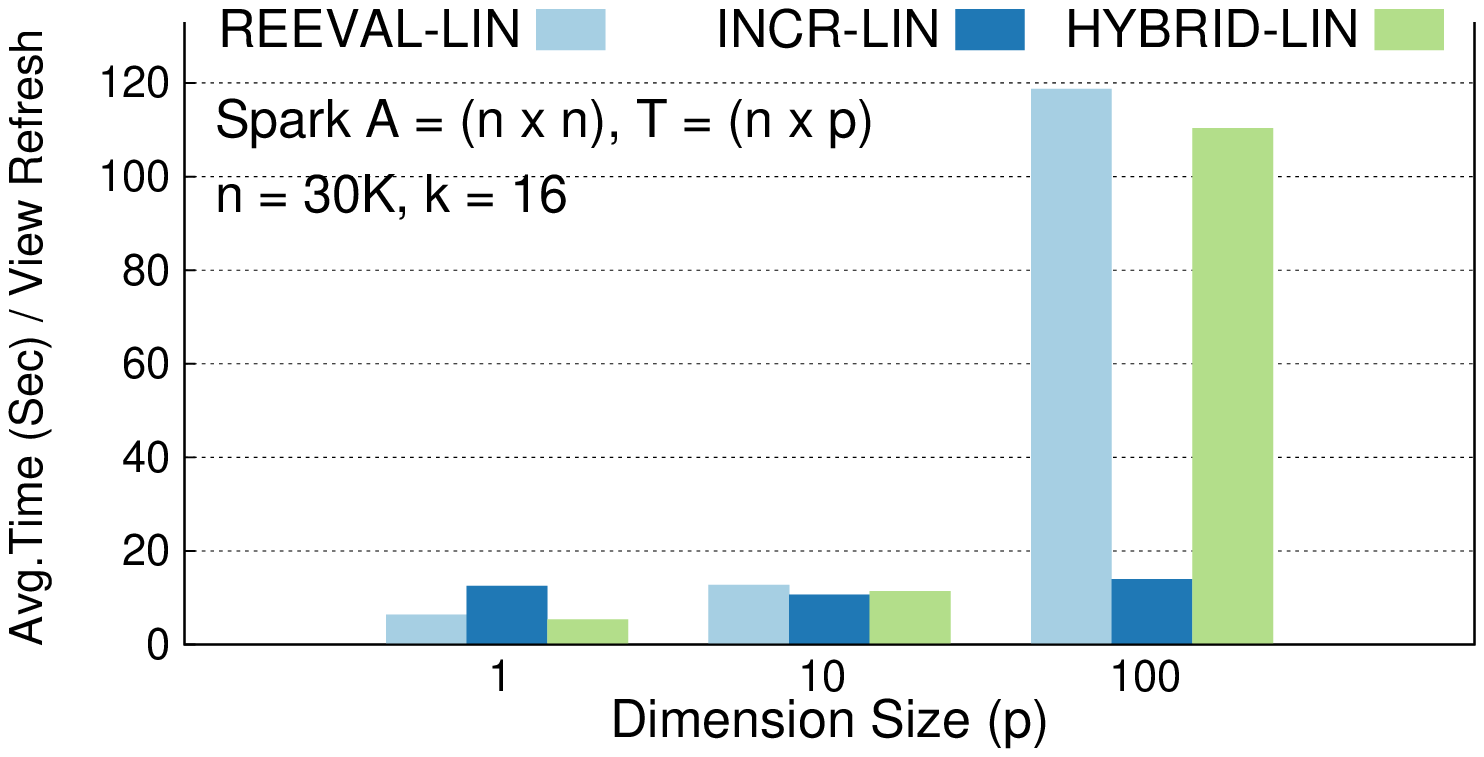}
                \vspace{-5mm}
                \caption{$T_{i+1}=AT_{i}$}
		            \label{fig:gen2}
        \end{subfigure}
        \begin{subfigure}[b]{0.245\textwidth}
                \centering
                \includegraphics[width=\textwidth]{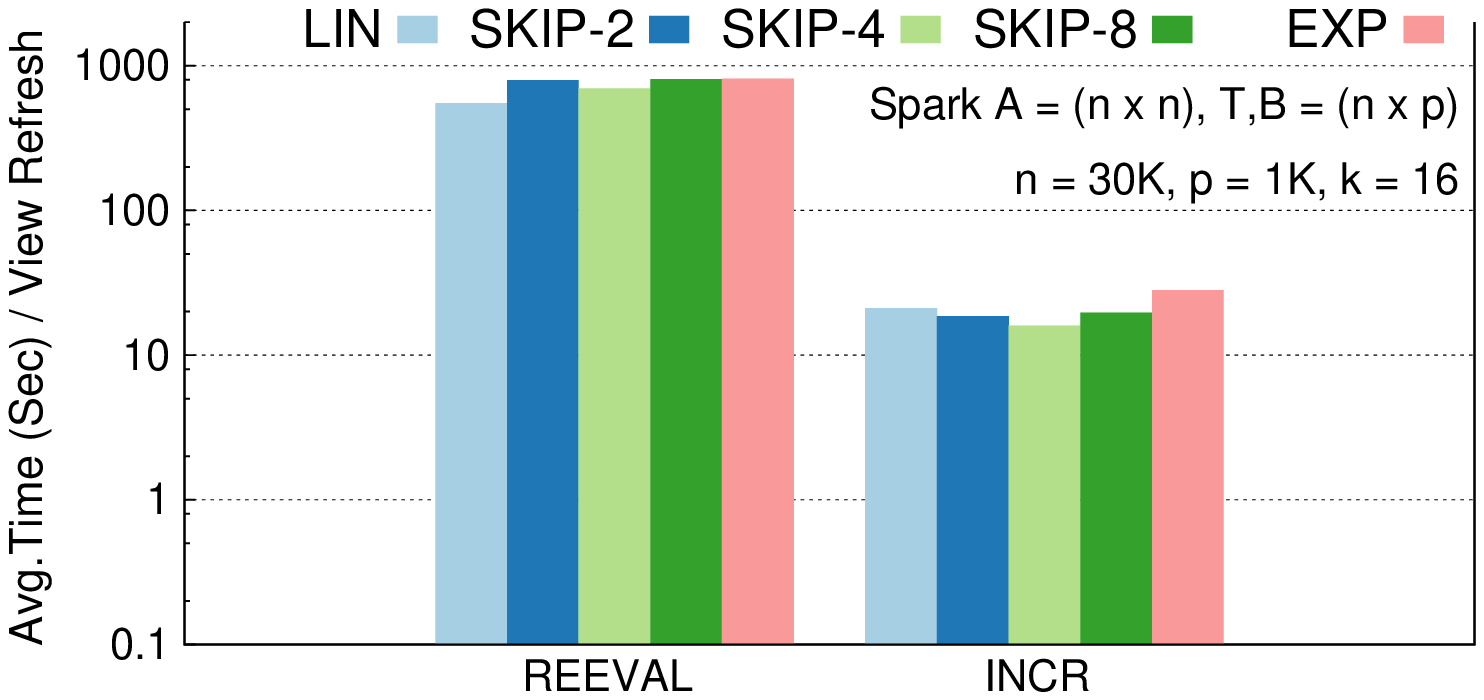}
                \vspace{-5mm} 
                \caption{LR: $T_{i+1}=AT_{i}+B$}
		            \label{fig:gen3}
        \end{subfigure}
\caption{Performance Evaluation of Incremental Maintenance using Octave and
Spark}
\end{figure*}

We have built the \system system that implements incremental maintenance of
analytical queries written as (iterative) linear algebra programs.
It is a compilation framework that transforms a given program, based on the
techniques discussed above, into efficient update triggers optimized for the
execution on different runtime engines.
Fig.~\ref{fig:system_overview} gives an overview of the system.

\noindent{\bf Workflow.}\ The \system framework consists of several compilation
stages:
the system transforms the code written in APL-style languages
(e.g., R, MATLAB, Octave) into an abstract syntax tree (AST), performs
incremental compilation, optimizes produced update triggers, and generates efficient code for execution
on single-node (e.g., MATLAB) or parallel processing platforms (e.g.,
Spark, Mahout, Hadoop).
The generated code consists of trigger functions for changes in each input
matrix used in the original program.

The optimizer analyzes intra- and inter-statement dependencies in the input
program and performs transformations, like common subexpression elimination and
copy propagation~\cite{Muchnick1997}, to reduce the overall maintenance cost.
In this process, the optimizer might define a number of auxiliary materialized
views that are maintained during runtime to support efficient processing of the
trigger functions.

\noindent{\bf Extensibility.}\ The \system framework is also extensible: one may add new frontends to transform
different input languages into AST or new backends that generate code for
various execution environments. 
At the moment, \system supports generation of Octave
programs that are optimized for execution in multiprocessor
environments, as well as Spark code for execution on large-scale
cluster platforms. The experimental section presents results for both backends.

%We envision \system being used as a back-end...

\noindent{\bf Distributed Execution.}\ 
The competitive advantage of incremental computation over re-evaluation 
-- reduced computation time -- is even more pronounced in distributed
environments.
Generated incremental programs are amenable to distributed execution as they are
composed of the standard matrix operations, for which many specialized tools
offer scalable implementations, like ScaLAPACK, Intel MKL, and
Mahout.
In addition, by transforming expensive matrix operations to work with smaller
datasets and representing changes in factored form, our incremental
techniques also minimize the communication cost as less data has to be shipped
over the network.

\noindent{\bf Data Partitioning.}\ \system analyzes data flow dependencies and data access
patterns in a generated incremental program to decide on a partitioning scheme
that minimizes data movement.
A frequently occurring expression in trigger programs is a multiplication of a
large matrix and a small delta matrix, typically performed in both directions
(e.g., $A \mul \Delta{A}$ and $\Delta{A} \mul A$ in Example~\ref{ex:simple_delta}).
To keep such computations strictly local, 
\system partitions large matrices both horizontally and vertically, that is,
each node contains one block of rows and one block of columns of a given matrix.
Although such a hybrid partitioning strategy doubles the memory consumption, it
allows the system to avoid expensive reshuffling of large
matrices, requiring only small delta vectors or low-rank matrices to be
communicated.

%Our focus in building \system has been on supporting scalable in-memory
%computations. As of future work, we plan to extend the system with the support
%for secondary storage. Similarly as with data partitioning, we plan to exploit
%data access patterns of an incremental program to construct a data layout
%strategy (e.g., for pages on a disk) that minimizes the I/O overhead.

\section{Experiments}
\label{sec:experiments}

This section demonstrates the potential of \system over traditional
re-evaluation techniques by comparing the average view refresh time for common
data mining programs under a continuous stream of updates.
We have built an APL-style frontend where users can provide their programs and
annotate dynamic matrices. The \system backend consists of two code generators
capable of producing Octave and Spark executable code, optimized for
the execution in multiprocessor and distributed environments. 
%In this section, we compare the performance of re-evaluation and incremental
%computation of a set of input queries.

For both Spark and Octave backends, our results show that:
\begin{inparaenum}[\itshape a\upshape)] \item Incremental view maintenance
outperforms traditional re-evaluation in almost all cases, validating the
complexity results of Section~\ref{sec:use_cases}; \item The performance gap
between re-evaluation and incremental computation increases with higher
dimensions; \item The hybrid evaluation strategy from
Section~\ref{subsec:general_form}, which combines re-evaluation and incremental
computation, exhibits best performance when the input matrices are not large
enough to justify the factored delta representation.
\end{inparaenum}

\noindent{\bf Experimental setup.}
To evaluate \system's performance using Octave,
we run experiments on a $2.66$GHz Intel Xeon with $2 \times 6$ cores, each with
$2$ hardware threads, $64$GB of DDR$3$ RAM, and Mac OS X Lion 10.7.5. We execute
the generated code using GNU Octave v$3.6.4$, an APL-style numerical computation framework, which
relies on the ATLAS library for performing multi-threaded BLAS
operations.

For large-scale experiments, we use an Amazon EC2
cluster with $26$ compute-optimized instances (c3.8xlarge). Each instance has
$32$ virtual CPUs, each of them is a hardware hyper-thread from a $2.8$GHz
Intel Xeon E5-2680v2 processor, $60$GB of RAM, and $2 \times 320$GB SSD.
The instances are placed inside a non-blocking $10$ Gigabit Ethernet network.
 
We run our experiments on top of the Spark engine -- an in-memory parallel
processing framework for large-scale data analysis. We configure Spark to launch
$4$ workers on one EC2 instance -- $100$ workers in total, each with $8$ virtual
CPUs and $13.6$GB of RAM -- and one master node on a separate EC2 instance. The
generated Spark code relies on Jblas for performing matrix
operations in Java. The Jblas library is essentially a wrapper around
the BLAS and LAPACK routines. For the purpose of our
experiments, we compiled Jblas with AMD Core Math Library v5.3.1 (ACML)
-- AMD's BLAS implementation optimized for high performance.
Jblas uses the ACML native library only for $\bigO(n^\gamma)$
operations, like matrix multiplication.

%To make use of the ACML native library, Jblas has to serialize Java
%arrays into native arrays, which might be an expensive operation. Thus,
% Jblas performs native calls only for $\bigO(n^\gamma)$ matrix operations, like
% matrix multiplication.

We implement matrix multiplication on top of Spark using the simple parallel
algorithm~\cite{IntroParallel:2003}, and we partition input matrices in a  $10
\times 10$ grid. For the scalability test we use square grids of smaller sizes.
For incremental evaluation, which involves multiplication with low-rank
matrices, we use the data partitioning scheme explained in
Section~\ref{sec:systems}; we split the data horizontally among all available nodes, then broadcast the
smaller relation to perform local computations, and finally
concatenate the result at the master node. The Spark framework carries out the
data shuffling among nodes.

\noindent{\bf Workload.}\ Our experiments consider dense random matrices up to 
$(100\mbox{K} \times 100\mbox{K})$ in size, containing up to $10$ billion
entries.
All matrices have double precision and are
preconditioned appropriately for numerical stability. For
incremental evaluation, we also precompute the initial values of all auxiliary
views and preload these values before the actual computation. We generate a
continuous random stream of rank-$1$ updates where each update affects one row
of an input matrix. On every such change, we re-evaluate or incrementally
maintain the final result. The reported values show the average view refresh
time over $3$ runs; the standard deviation was less than $5$\% in each
experiment.

\noindent{\bf Notation.}\ Throughout the evaluation discussion we use the following notation:
\begin{inparaenum}[\itshape a\upshape)]
\item The prefixes \textsc{Reeval}, \textsc{Incr}, and
\textsc{Hybrid} denote traditional re-evaluation, incremental processing, and
hybrid computation, respectively;
\item The suffixes \textsc{Lin}, \textsc{Exp}, and \textsc{Skip-s}
represent the linear, exponential, and skip-$s$ models, respectively.
\end{inparaenum}These evaluation models are described in detail in Section~\ref{sec:use_cases} and Table~\ref{table:time_space_cost}.

\noindent{\bf Ordinary Least Squares.}\ We conduct a set of experiments to
evaluate the statistical estimator $\beta^*$ using OLS as defined in
Section~\ref{subsec:ols}.  
Exceptionally in this example, we present only the Octave results
as the current Spark backend lacks the support for re-evaluation of
matrix inversion. The predictors matrix $X$ has dimension $(n \times n)$
and the responses matrix $Y$ is of dimension $(n \times p)$. Given a continuous
stream of updates on $X$, Fig.~\ref{fig:ols} compares the average execution time
of re-evaluation \textsc{Reeval} and incremental maintenance \textsc{Incr} of
$\beta^*$ with different sizes of $n$. We set $p = 1$ because
this setting represents the lowest cost for \textsc{Reeval}, as the cost is
dominated by the matrix inversion re-evaluation $\bigO(n^\gamma)$. The graph
illustrates the superiority of \textsc{Incr} over \textsc{Reeval} in computing
the OLS estimates.
Notice the asymptotically different behavior of these two graphs -- the performance gap between \textsc{Reeval} and
\textsc{Incr} increases with matrix size, from $3.56$x for $n=4,000$ to
$11.45$x for $n=20,000$. This is consistent with the complexity results
from Section \ref{subsec:ols}.

\noindent{\bf Matrix Powers.}\ We analyze the performance of 
the matrix powers $A^k$ evaluation, where $A$ has dimension $(n \times n)$, by
varying different parameters of the computational model. First, we evaluate the performance of the evaluation 
strategies presented in Section~\ref{subsec:matrix_powers}, for a fixed
dimension size and number of iterations $k = 16$. Fig.~\ref{fig:mp1} illustrates 
the average view refresh time of Octave generated programs for $n = 10,000$ and
Spark generated programs for $n = 30,000$. In both implementations, the results 
demonstrate the virtue of \textsc{Incr} over \textsc{Reeval} and the efficiency 
of \incrExp over \incrLin and \incrSkips{s}.

Next, we explore various scalability aspects of the matrix powers computation. 
Fig.~\ref{fig:mp2} reports on the Octave performance over larger
dimension sizes $n$, given a fixed number of iteration steps $k=16$;
Fig.~\ref{fig:mp2} also illustrates the Spark performance for even
larger matrices. In both cases, \incrExp outperforms
\reevalExp with similar asymptotic behavior. As in the OLS example,
the performance gap increases with higher dimensionality.

Fig.~\ref{fig:mp2} shows the Spark re-evaluation results
for matrices up to size $n=50,000$. Beyond this limit, the running time of
re-evaluation exceeds one hour due to an increased communication
cost and garbage collection time. The re-evaluation strategy has a more
dynamic model of memory usage due to frequent allocation and deallocation of
large memory chunks as the data gets shuffled among nodes. 
%Also, the Spark framework writes all shuffled data to disks for recoverability.
In contrast, incremental evaluation avoids expensive communication by sending
over the network only relatively small matrices. Up until $n=90,000$, we see
a linear increase in the \incrExp running time. However, as discussed in Section
\ref{sec:use_cases}, we expect the $\bigO{(n^2)}$
complexity for incremental evaluation. The explanation lies in that the generated Spark code
distributes the matrix-vector computation among many nodes and,
inside each node, over multiple available cores, effectively achieving linear
scalability.
For $n=100,000$, incremental evaluation hits the resource limit in
this cluster configuration, causing garbage collection to increase the average
view refresh time.

\begin{table}[t]
\begin{center}
{\renewcommand{\arraystretch}{1.2}    
  \scriptsize
  \begin{tabular}{l@{\phantom{abcd}}l@{\phantom{abcd}}r@{\phantom{abcd}}r@{\phantom{abcd}}r@{\phantom{abcd}}r}
  \toprule
%   &  & \multicolumn{4}{c}{\textbf{Matrix Size}} \\
%  \cmidrule{3-6} 
  \multicolumn{2}{l}{\textbf{Matrix Size}} & $20$K & $30$K & $40$K & $50$K \\
  \midrule
  Memory & \reevalExp & $8.9$ & $20.1$ &
  $35.8$ & $55.9$
  \\
  (GB) & \incrExp & $29.8$ & $67.1$ & $119.2$ & $186.3$ \\
\midrule
 Time & \reevalExp & $95.0$ & $203.4$ &
 $667.3$ & $1328.7$ \\
 (sec) & \incrExp & $9.6$ & $14.1$ & $21.0$ & $24.9$ \\
\midrule
 \multicolumn{2}{l}{Speedup vs. Memory Cost} & $2.99$ & $4.31$ & $9.55$ &
 $16.00$ \\
\bottomrule
\end{tabular}}
\end{center}
\vspace{-3ex}
\caption{The memory requirements and Spark view refresh times of
\reevalExp and \incrExp for $A^{16}$ and different matrix sizes. The
last row is the ratio between the speedup and memory overhead incurred by
maintaining auxiliary views.}
\vspace{-4ex}
\label{table:time_memory_reqs}
\end{table}

\noindent{\bf Memory Requirements.}\ Table~\ref{table:time_memory_reqs} presents
the memory requirements and Spark single-update execution times of \reevalExp
and \incrExp for the $A^{16}$ computation and various matrix dimensions. The
last row represents the ratio between the speedup achieved using incremental
evaluation and the memory overhead imposed by maintaining the results of intermediate iterations. We conclude
that the benefit of investing more memory resources increases with higher
dimensionality of the computation.

Next, we evaluate the scalability of the matrix powers computation for
different numbers of Spark nodes. We evaluate various square grid
configurations for re-evaluation of $A^{16}$, where
$n=30,000$\footnote{\scriptsize To achieve perfect load balance with different
grid configurations, we choose the matrix size to be the closest number to $30,000$ that is divisible by the total
number of workers.}.
Fig.~\ref{fig:mp_scalability_nodes} shows that our Spark implementation of
matrix multiplication scales with more nodes. Also note that incremental
evaluation is less susceptible to the number of nodes than re-evaluation; the 
average time per view refresh varies from 10 to 26 seconds.

Finally, in Fig.~\ref{fig:mp3}, we vary the number of iteration steps $k$ 
given a fixed dimension, $n=10,000$ for Octave and $n=30,000$ for Spark. The
Octave performance gap between \incrExp and \reevalExp increases with more
iterations up to $k=256$ when the size of the delta vectors $(10,000 \times
256)$ becomes comparable with the matrix size. The Spark implementation
broadcasts these delta vectors to each worker, so the achieved speedups
decrease with larger iteration numbers due to the increased communication costs.
However, as argued in Section~\ref{subsec:matrix_powers}, many iterative
algorithms in practice require only a few iterations to converge, and for those 
the communication costs stay low.

\noindent{\bf Sums of Powers.}\ We analyze the computation of sums
of matrix powers, as described in Section~\ref{subsec:sums_matrix_powers}. Since
it shares the same complexity as the matrix powers computation, we present only 
the performance of the exponential models.
Fig.~\ref{fig:Sums} compares \incrExp and \reevalExp on various dimension sizes
$n$ using Octave and Spark, for a given fixed number of iterations $k=16$.
Similarly to the matrix powers results from Fig.~\ref{fig:mp2}, \incrExp 
outperforms traditional \reevalExp, and the achieved speedup increases with $n$.
Beyond $n=40,000$, the Spark re-evaluation exceeds the one-hour time limit.

\noindent{\bf General Form.}\ We evaluate the general
iterative model of computation $T_{i+1} = A \mul T_{i} + B$, where
$T_{n \times p}$, $A_{n \times n}$, and $B_{n \times p}$, using the following settings (due to
space constraints we show only important Spark results):
% and omit Octave evaluation):
%; see our technical report for more detailed analysis):

\smartparagraph{$\bullet$ B=0.}{ The iterative computation degenerates to
$T_{i+1} = A \mul T_{i}$, which represents matrix powers when $p=n$, and thus we
explore an alternative setting of $1\leq p<n$. For small values of $p$, the
\textsc{Lin} model has the lowest complexity as it avoids expensive
$\bigO{(n^\gamma)}$ matrix multiplications. Fig.~\ref{fig:gen2} shows the
results of different evaluation strategies, given a fixed dimension $n=30,000$
and iteration steps $k=16$. For $p=1$, \textsc{HybridLin}
outperforms \textsc{ReevalLin} by $16\%$ and \textsc{IncrLin} by $53\%$.
However, the evaluation cost of both \textsc{HybridLin} and \textsc{ReevalLin}
increases linearly with $p$. \textsc{IncrLin} exhibits the best performance
among them when $p$ is large enough to justify the factored delta
representation.}

\smartparagraph{$\bullet$ B$\neq$0.}{ We study an analytical query evaluating 
linear regression using the gradient descent algorithm of the form 
$\Theta_{i+1} = \Theta_{i}- X^{T}(X\Theta_{i}-Y)$.
We adapt this form to the general iterative model by substituting 
$A=I-X^{T}X$ and $B=X^{T}Y$, where $I$ represents the identity matrix.
Fig.~\ref{fig:gen3} shows the performance of different iterative models for
both re-evaluation and incremental computation, given fixed sizes $n = 30,000$
and $p = 1,000$ and a fixed number of iterations $k=16$. Note the
logarithmic scale on the $y$ axis. The \textsc{Lin} model exhibits the best
re-evaluation performance; the \textsc{Skip-4} model has the lowest view refresh
time for incremental evaluation. Overall, incremental computation outperforms
traditional re-evaluation by a factor of 36.7x.
}

\begin{table}[t]
\begin{center}
{\renewcommand{\arraystretch}{1.2}    
  \scriptsize
  \begin{tabular}{l@{\phantom{ab}}c@{\phantom{ab}}c@{\phantom{ab}}c@{\phantom{ab}}c@{\phantom{ab}}c@{\phantom{ab}}c@{\phantom{ab}}c}
  \toprule
%   &  & \multicolumn{4}{c}{\textbf{Matrix Size}} \\
%  \cmidrule{3-6} 
  \textbf{Zipf factor} & $5.0$ & $4.0$ & $3.0$ & $2.0$ & $1.0$ & $0.0$
  \\
  \midrule
  Octave (10K) & $6.3$ & $6.8$ & $7.5$ & $10.9$ & $68.4$ & $236.5$
  \\
  Spark (30K) & $28.1$ & $41.5$ & $67.3$ & $186.1$ & $508.9$ & $1678.8$ \\
\bottomrule
\end{tabular}}
\end{center}
\vspace{-3ex}
\caption{The average Octave and Spark view refresh times in seconds for
\incrExp of $A^{16}$ and a batch of $1,000$ updates. The row
update frequency is drawn from a Zipf distribution.}
\vspace{-4ex}
\label{table:batch_updates}
\end{table}

\noindent{\bf Batch updates.}\ We analyze the performance of
incremental matrix powers computation for batch updates.
We simulate a use case in which certain regions of the input matrix are changed
more frequently than the others, and the frequency of row updates is
described using a Zipf distribution. 
Table \ref{table:batch_updates} shows the performance of incremental
evaluation for a batch of $1,000$ updates and different Zipf factors. As the
row update frequency becomes more uniform, that is, more rows are affected by a
given batch, \incrExp loses its advantage over \reevalExp because the delta
matrices become larger and more expensive to compute and distribute. To put
these results in the context, a single update of a $n=10,000$ matrix in
Octave takes $99.1$ and $6.3$ seconds on average for \reevalExp and \incrExp; 
For one update of a $n=30,000$ matrix using Spark, \reevalExp and \incrExp take
$203.4$ and $14.1$ seconds on average. We observe that the Spark implementation 
exhibits huge communication overhead, which significantly prolongs the running 
time. 
We plan to investigate this issue in our future work.
%\vspace{-2.0mm}

{
%\vspace{0.5cm}
\scriptsize
%\footnotesize
\bibliographystyle{abbrv}
\bibliography{references}
}

\normalsize
\appendix

\section{Matrix Powers}
\label{sec:appendix_matrix_powers}

We analyze the time complexity of incremental view maintenance and hybrid
evaluation for the computation of matrix powers discussed in
Section~\ref{sec:use_cases} for rank-$1$ updates to $A$, denoted by $\Delta{A} =
\du \mul \tr{\dv}$.
The analysis considers the three iterative models of computation from
Section~\ref{sec:linear_programs}.

\smartparagraph{$\bullet$}{ \ In the linear model, $\Delta{P_1} = \du \mul
\tr{\dv}$, thus $\dU_1 = \du$ and $\dV_1 = \dv$. For $i > 1$, 
\vspace{-1mm}
\begin{align*}
    \Delta{P_i} = 
      \begin{bmatrix}
        \dU_1 & (A \mul \dU_{i-1} + \dU_1 \mul (\tr{\dV_1} \mul \dU_{i-1}))
      \end{bmatrix} \mul
      \begin{bmatrix}
        \tr{\dV_1} \mul P_{i-1} \\
        \tr{\dV_{i-1}}
      \end{bmatrix}      
\end{align*}
where $\dU_i$ and $\dV_i$ are $(n \times i)$ block matrices
(provable by induction), and $A$ and $P_i$ are $(n \times n)$ matrices.
The evaluation cost of $P_i$, when $i > 1$, is:
\begin{align*}
  cost_{\textsc{Lin}}(P_i) &= cost(\dU_i) + cost(\dV_i) + cost(\dU_i \mul
  \tr{\dV_i}) \\
%  &= (i-1)n^2 + (i-1)n + (i-1)n + (i-1)n + n^2 + in^2 \\
  &= 2n^2i + 3n(i-1).
\end{align*}
%
%\vspace{-2mm}
\noindent The evaluation cost of all iterations in program $\mathcal{P}$ is
\vspace{-1mm}
\begin{align*}
  cost_{\textsc{Lin}}(\mathcal{P}) &= n^2 + \sum_{i=2}^k (2n^2i + 3n(i-1)) \\
  &= n^2(k^2 + k - 1) + \frac{3}{2}nk(k-1) = \bigO(n^2k^2).
\end{align*}
}

\vspace{-1mm}
\smartparagraph{$\bullet$}{ \ In the exponential model, $\Delta{P_1} = \du \mul
\tr{\dv}$, thus $\dU_1 = \du$ and $\dV_1 = \dv$. For $i > 1$,
\vspace{-1mm}
\begin{align*}
   \Delta{P_i} = 
       \begin{bmatrix}
          \dU_{i/2} & (P_{i/2} \mul \dU_{i/2} + \dU_{i/2} \mul (\tr{\dV_{i/2}} \mul \dU_{i/2}))
       \end{bmatrix} \mul
       \begin{bmatrix}
          \tr{\dV_{i/2}} \mul P_{i/2} \\ \tr{\dV_{i/2}}
       \end{bmatrix} 
\end{align*}
where $\dU_i$ and $\dV_i$ are $(n \times i)$ block matrices, and $P_i$ is an $(n
\times n)$ matrix.
The evaluation cost of $P_i$, when $i > 1$, is
%\vspace{-1mm}
\begin{align*}
  cost_{\textsc{Exp}}(P_i) &= cost(\dU_i) + cost(\dV_i) + cost(\dU_i \mul
  \tr{\dV_i}) \\
%  &= \frac{i}{2}n^2 + \frac{i}{2}n + \left(\frac{i}{2}\right)^2 n + 
%     \left(\frac{i}{2}\right)^2 n + \frac{i}{2}n^2 + in^2 \\      
  &= 2n^2i + n\left(\frac{i^2}{2} + \frac{i}{2}\right).
\end{align*}
The evaluation cost of all iterations in program $\mathcal{P}$ is:
\vspace{-1mm}
\begin{align*}
  cost_{\textsc{Exp}}(\mathcal{P}) &= n^2 + \sum_{i=2,4,8, \ldots k} (2n^2i +
  n\frac{i^2}{2} + n\frac{i}{2}) \\
  &= n^2(4k-3) + n\frac{(k-1)(2k+5)}{3} = \bigO(n^2k).
\end{align*}
}

\vspace{-1mm}
\smartparagraph{$\bullet$}{ \ In the skip model, we first evaluate $T_s$ using
the exponential model in $\bigO(n^2s)$ time as discussed above.  Then
$\Delta{P_s} = \dU_s \mul \tr{\dV_s}$, where $\dU_s$ and $\dV_s$ are  $(n \times
s)$ block matrices. For $i > s$, 
\vspace{-1mm}
\begin{align*}
  \Delta{P_i} =
       \begin{bmatrix}
          \dU_{s} & (P_{s} \mul \dU_{i-s} + \dU_{s} \mul (\tr{\dV_{s}} \mul \dU_{i-s}))
       \end{bmatrix} \mul
       \begin{bmatrix}
          \tr{\dV_{s}} \mul P_{i-s} \\ \tr{\dV_{i-s}}
       \end{bmatrix} 
\end{align*}
where $\dU_i$ and $\dV_i$ are $(n \times i)$ block matrices, and $P_i$ is an $(n
\times n)$ matrix.
The evaluation cost of $P_i$, when $i > s$, is
\vspace{-1mm}
\begin{align*}
  cost_{\textsc{Skip}}(P_i) &= cost(\dU_i) + cost(\dV_i) + cost(\dU_i \mul
  \tr{V_i}) \\
%  = (i-s)n^2 + n(i-s) + s(i-s)n + s(i-s)n + sn^2 + in^2 \\
  &= 2n^2i + n(2s+1)(i-s).
\end{align*}
The evaluation cost of all iterations in program $\mathcal{P}$ is
\vspace{-1mm}
\begin{align*}
  cost_{\textsc{Skip}}(\mathcal{P}) &= \bigO(n^2s) + 
      \sum_{i=2s,3s,\ldots k} (2n^2i + n(2s+1)(i-s)) \\
%  cost_{\textsc{Skip}}(P) &= (4s-3)n^2 + \frac{(s-1)(2s+5)}{3}n + 
%      \sum_{i=2s,3s,\ldots k} (2in^2 + (2s+1)(i-s)n) \\
%  = \left(\frac{k^2}{s} + k + 2s - 3 \right)n^2 + 
%     \left( \frac{(s-1)(2s+5)}{3} + \frac{(2s+1)k\left(\frac{k}{s} - 1\right)}{2} \right)n
  &= \bigO(n^2s) + \bigO\left(n^2\frac{k^2}{s}\right) =
  \bigO\left(n^2\frac{k^2}{s}\right).
\end{align*}
}

\vspace{-3mm}
\section{General Form: T\textsubscript{\bf{i+1}} = AT\textsubscript{\bf{i}} + B}
\label{sec:appendix_atb}
We analyze the time complexity of incremental view maintenance for the
computation of matrix powers discussed in Section~\ref{sec:use_cases} for
rank-$1$ updates to $A$, denoted by $\Delta{A} = \du \mul \tr{\dv}$.
The analysis considers the three iterative models of computation from
Section~\ref{sec:linear_programs}.

% {\em Detailed Analysis.}
The incremental strategy maintains the result of $T_i$, and if needed $P_i$ and
$S_i$, for updates to $A$, denoted by $\Delta{A} = \du \mul \tr{\dv}$. The delta
expressions are in the factored form as $\Delta{T_i} = \dU_i \mul \dV_i$,
$\Delta{P_i} = \dQ_i \mul \tr{\dR_i}$, and $\Delta{S_i} = \dZ_i \mul
\tr{\dW_i}$.

\smartparagraph{$\bullet$}{ \ In the linear model, $\Delta{A} = \du \mul
\tr{\dv}$, thus $\dU_1 = \du$ and $\dV_1 =  \tr{T_0} \mul \dv$. For $i > 1$,
\vspace{-1mm}
\begin{align*}
    \Delta{T_i} = 
          \begin{bmatrix}
            \du & (A \mul \dU_{i-1} + \du \mul (\tr{\dv} \mul \dU_{i-1}))
          \end{bmatrix} \mul
          \begin{bmatrix}
            \tr{\dv} \mul T_{i-1} \\
            \tr{\dV_{i-1}}
          \end{bmatrix} 
\end{align*}
where $\dU_{i} = (n \times i)$ and $\dV_{i} = (p \times i)$ (provable by induction).
 %, $A = (n \times n)$, and $T_i = (n \times p)$.
%
The evaluation cost of $T_i$, when $i > 1$, is
\vspace{-1mm}
\begin{align*}
  cost_{\textsc{Lin}}(T_i) &= cost(\dU_i) + cost(\dV_i) + cost(\dU_i \mul
  \tr{\dV_i}) \\
%  &= (i-1)n^2 + (i-1)n + (i-1)n + (i-1)n + np + inp \\
  &= n^2(i-1) + 3n(i-1) + np(i+1).
\end{align*}
The evaluation cost of all iterations in program $\mathcal{P}$ is
\vspace{-1mm}
\begin{align*}
  cost_{\textsc{Lin}}(\mathcal{P}) &= 2np + \sum_{i=2}^k (n^2(i-1) + np(i+1) +
  \bigO(ni)) \\
  &= \bigO((n^2 + np)k^2).
\end{align*}
}

\vspace{-1mm}
\smartparagraph{$\bullet$}{ \ In the exponential model, $\Delta{A} = \du \mul
\tr{\dv}$, thus $\dU_1 = \du$ and $\dV_1 =  \tr{T_0} \mul \dv$.
Let $\Delta{P_i} = \dQ_i \mul \tr{\dR_i}$ and $\Delta{S_i} = \dZ_i \mul \tr{\dW_i}$.
For $i > 1$,
\begin{align*}
   &\dU_i = 
       \begin{bmatrix}
          \dQ_{i/2} & 
          (P_{i/2} \mul \dU_{i/2} + \dQ_{i/2} \mul (\tr{\dR_{i/2}} \mul \dU_{i/2})) & 
          \dZ_{i/2}
       \end{bmatrix} \\
   &\dV_i =        
       \begin{bmatrix}
          \tr{T_{i/2}} \mul \dR_{i/2} & 
          \dV_{i/2} &
          \tr{B} \mul dW_{i/2} 
       \end{bmatrix}        
%   \Delta{P_i} = 
%       \begin{bmatrix}
%          \dQ_{i/2} & (P_{i/2} \mul \dU_{i/2} + \dQ_{i/2} \mul (\tr{\dR_{i/2}} \mul \dU_{i/2})) & \dZ_{i/2}
%       \end{bmatrix} \mul
%       \begin{bmatrix}
%          \tr{\dR_{i/2}} \mul T_{i/2} \\ \tr{\dV_{i/2}} \\ \tr{\dW_{i/2}} \mul B
%       \end{bmatrix} 
\end{align*}
where $\dU_i = (n \times (2i-1))$ and $\dV_i = (p \times (2i-1))$.
%, $\dQ_i = (n \times i)$, $\dR_i = (n \times i)$, $\dZ_i = (n \times i)$, $\dW_i = (n \times i)$, $P = (n \times n)$.
%   
The evaluation cost of $T_i$, when $i > 1$, disregarding the computation of $P_i$ and $S_i$, is
\begin{align*}
  cost_{\textsc{Exp}}(T_i) &= cost(\dU_i) + cost(\dV_i) + cost(\dU_i \mul
  \tr{\dV_i}) \\
  &= n^2(i-1) + np(3i-1) + n(i+1)(i-1).
\end{align*}
The evaluation cost of all iterations in program $\mathcal{P}$ is
\begin{align*}
  cost_{\textsc{Exp}}(\mathcal{P}) =& \bigO(kn^2) + \\ 
  & \sum_{i=2,4, \ldots k} \!\!\!\!\! (n^2(i-1) + np(3i-1) + \bigO(ni^2)) \\
  =& \bigO((n^2 + np)k).
\end{align*}
}

\vspace{-1mm}
\smartparagraph{$\bullet$}{ \ In the skip model, we first evaluate $T_s$ using
the exponential model in $\bigO(s(n^2+np))$ time as discussed above. Then 
$\Delta{T_s} = \dU_s \mul \tr{\dV_s}$, where $\dU_s$ and $\dV_s$ are $(n \times
(2s-1))$ block matrices. For $i > s$,
\begin{align*}
   \Delta{T_i} = 
       \begin{bmatrix}
          \dQ_{s} & (P_{s} \mul \dU_{i-s} + \dQ_{s} \mul (\tr{\dR_{s}} \mul \dU_{i-s})) & \dZ_{s}
       \end{bmatrix} \mul
       \begin{bmatrix}
          \tr{\dR_{s}} \mul T_{i-s} \\ \tr{\dV_{i-s}} \\ \tr{\dW_{s}} \mul B
       \end{bmatrix} 
\end{align*}
where $\dU_i$ and $\dV_i$ are $(n \times (2i-1))$ block matrices.
The evaluation cost of $T_i$, when $i > s$, disregarding the computation of $P_i$ and $S_i$, is
\begin{align*}
  cost_{\textsc{Skip}}(T_i) &= cost(\dU_i) + cost(\dV_i) + cost(\dU_i \mul
  \tr{V_i}) \\
%  = (2(i-s)-1)n^2 + (2(i-s)-1)n + 2(2(i-s)-1)ns + 2nps + np(2i-1)
  &= (2n^2(i-s)-1) + np(2i-1) + \bigO(n).
\end{align*}
The evaluation cost of all iterations in program $\mathcal{P}$ is
\begin{align*}
  cost_{\textsc{Skip}}(\mathcal{P}) =& \bigO((n^2 + np)\log{s}) + \\
    & \sum_{i=2s,3s,\ldots k} \!\!\!\!\!(n^2(2(i-s)-1) + np(2i-1) + \bigO(n)) \\
  =& \bigO((n^2 + np)\log{s}) + \bigO\left((n^2 + np)\frac{k^2}{s}\right) \\
  =& \mathcal{O}\left((n^2 + np)\frac{k^2}{s}\right).
\end{align*}
}

%\smallskip

We analyze the cost of the hybrid evaluation strategy.

\smartparagraph{$\bullet$}{ \ In the linear model,  $\Delta{A} = \du \mul
\tr{\dv}$ and $\Delta{T_1} = \Delta{A} \mul T_0$. For $i > 1$, 
\vspace{-1mm}
\begin{align*}
  \Delta{T_i} 
    = \du \mul \tr{\dv} \mul T_{i-1} + A \mul \Delta{T_{i-1}} + \du \mul \tr{\dv} \mul \Delta{T_{i-1}}.
\end{align*}
The evaluation cost of $T_i$, when $i > s$, is
%\vspace{-1mm}
\begin{align*}
  cost_{\textsc{Lin}}(T_i) &= cost(\Delta{T_i}) 
%    = 2np + np + pn^2 + np + 2np \\
    = pn^2 + 6np.
\end{align*}
The evaluation cost of all iterations in program $\mathcal{P}$ is
\vspace{-1mm}
\begin{align*}
  cost_{\textsc{Lin}}(\mathcal{P}) = n^2 + \sum_{i=2}^k (pn^2 + 6np) 
  = \bigO(pn^2k).
\end{align*}
}

\vspace{-1mm}
\smartparagraph{$\bullet$}{ \ In the exponential model, $\Delta{A} = \du \mul
\tr{\dv}$ and $\Delta{T_1} = \Delta{A} \mul T_0$.
Let $\Delta{P_i} = \dQ_i \mul \tr{\dR_i}$ and $\Delta{S_i} = \dZ_i \mul \tr{\dW_i}$.
For $i > 1$,
\begin{align*}
  \Delta{T_i} 
    =& \dQ_{i/2} \mul \tr{\dR_{i/2}} \mul T_{i/2} + P_{i/2} \mul \Delta{T_{i/2}} 
    + \\
    & \dQ_{i/2} \mul \tr{\dR_{i/2}} \mul \Delta{T_{i/2}} + \dZ_{i/2} \mul
    \tr{\dW_{i/2}} \mul B.
\end{align*}
The evaluation cost of $T_i$, when $i > 1$, disregarding the computation of $P_i$ and $S_i$, is
\vspace{-1mm}
\begin{align*}
  cost_{\textsc{Exp}}(T_i) = cost(\Delta{T_i}) 
%  = \frac{i}{2}np + \frac{i}{2}np + np + pn^2 + pn + \frac{i}{2}np + \frac{i}{2}np + np + \frac{i}{2}np + \frac{i}{2}np
  = pn^2 + 3np(i+1).
\end{align*}
The evaluation cost of all iterations in program $\mathcal{P}$ is
\vspace{-1mm}
\begin{align*}
  cost_{\textsc{Exp}}(\mathcal{P}) &= \bigO(n^2k) + \sum_{i=2,4,8, \ldots k}
  (pn^2 + 3np(i+1)) \\
  &= \bigO(pn^2\log{k} + n^2k).
\end{align*}
}

\vspace{-1mm}
\smartparagraph{$\bullet$}{ \ In the skip model, we first evaluate $T_s$ using
the exponential model in $\bigO(pn^2\log{s} + sn^2)$ time as discussed above. 
For $i > s$, 
\begin{align*}
  \Delta{T_i} 
    = \dQ_s \mul \tr{\dR_s} \mul T_{i-s} + P_{s} \mul \Delta{T_{i-s}} + \dQ_s \mul \tr{\dR_s} \mul \Delta{T_{i-s}} + \dZ_{s} \mul \tr{\dW_{s}} \mul B.
\end{align*}
The evaluation cost of $T_i$, when $i > s$, disregarding the computation of 
$P_i$ and $S_i$, is 
\begin{align*}
  cost_{\textsc{Skip}}(T_i) = cost(\Delta{T_i}) 
%  &= snp + snp + np + pn^2 + np + snp + snp + np + snp + snp 
  = pn^2 + 6nps + 3np.
\end{align*}
The evaluation cost of all iterations in program $\mathcal{P}$ is
\begin{align*}
  cost_{\textsc{Skip}}(\mathcal{P}) =& \bigO(pn^2\log{s} + n^2s) + \\
  & \sum_{i=2s,3s,\ldots k} (pn^2 + 6nps + 3np) \\
  =& \bigO(pn^2\log{s} + n^2s) + \bigO\left(pn^2\frac{k}{s}\right) \\
  =& \bigO\left(pn^2\left(\log{s} + \frac{k}{s}\right) + n^2s\right).
\end{align*}
}

\end{document}